\documentclass[12pt,preprint]{aastex}
\newcommand{\kms}{{\,\rm km\,s}^{-1}} 
\newcommand{\ksm}{{\,\rm km}\ {\rm~s}^{-1}\ {\rm~Mpc}^{-1}}

\def\la{\mathrel{\hbox{\rlap{\hbox{\lower4pt\hbox{$\sim$}}}\hbox{$<$}}}}
\def\sun{\hbox{$\odot$}}
\renewcommand{\mag}{\mbox{$\;$mag}}

\begin{document}
\title{
     Bias Properties of Extragalactic Distance Indicators XII:
     Bias Effects of Slope Differences and Intrinsic Dispersion 
     on Tully-Fisher Distances to Galaxy Clusters with
     Application to the Virgo Cluster}
\author{Allan Sandage}
\affil{The Observatories of the Carnegie Institution of Washington,\\
       813 Santa Barbara Street, Pasadena, CA 91101-1292}
%
%
\affil{What seems so simple is often highly complex in its dreadful
    detail:\\ {\rm Nilo Negge in the Book of Wisdom, 1619}}

\begin{abstract}
The Teerikorpi incompleteness bias in the distance modulus 
of a galaxy cluster that is determined from incomplete data using 
the Tully-Fisher (TF) method is discussed differently than has 
been done in earlier papers of this series. A toy cluster is made 
with zero intrinsic TF dispersion but with slopes that differ 
between the calibrators and the cluster data, showing the bias 
caused by incorrect slopes. Intrinsic dispersion is added to the 
model and two strategies are used to analyze the data; first by 
binning the data by line width and then by apparent magnitude 
(the direct method), and second by binning by magnitude and then 
summing over all line widths (the inverse method). To illustrate 
these strategies, a composite cluster is made by combining the 
observations of Virgo~A and B subclusters with those for the Ursa 
Major I and II clusters, corrected to the Virgo~A distance. The 
cluster data are calibrated using Cepheid distances to 25 
galaxies that have adequate TF properties. Different moduli 
calculated with varying completeness limits are displayed. The 
cluster modulus derived from the complete cluster sample gives    
$(m - M)^{0} = 31.42\pm0.2$ (external) for Virgo~A, $31.80\pm0.16$
for Virgo~B, $31.26\pm0.13$ for UMa~I, and $31.58\pm0.17$ for UMa~II. 
Combining the Virgo~A distance ($D = 19.2\;$Mpc that has a range 
from 17.5 to 21.1$\;$Mpc) with its expansion velocity of 
$1175\pm50\kms$ as tied to the remote kinematic frame gives a Hubble 
constant of $61\ksm$ with a range from 53 to 70. 
\end{abstract}
\keywords{Research --- Tutorial --- Galaxies --- Data Analysis and Techniques}

\section{INTRODUCTION}
\label{sec:01}
The literature on the effect of bias in the measurement of 
extragalactic distances is extensive. Taken is a whole it seems  
comprehensive and exhaustive. Why then propose again an expos{\'e} 
of parts of the subject? The problem is not so simple in 
different situations, and new illustrations of the details can be 
useful.

     A significant review, complete to 1996, is by 
\citet{Teerikorpi:97}. He treats many aspects of the problem beginning
with the early insights of \citet{Kapteyn:14}, and then proceeds to
the classical \citet{Malmquist:20,Malmquist:22} and \citet{Scott:57}
effects.  

     The Malmquist calculation only concerns the bias integrated 
over a total sample. It does not show how the biased distance 
errors change as the truncation level is made fainter. Therefore, 
the classical Malmquist bias is not useful in correcting 
individual distances at various magnitude levels above the 
truncation level. For that, more complicated correction 
procedures mentioned below, are needed.

     An equally complicated problem is to analyze the bias effects 
for individual galaxy distances using the Tully-Fisher line width 
(LW) method applied to field galaxies chosen in different ways, 
such as from magnitude-limited catalogs, or by imposing apparent 
diameter limits on 21-cm lists before measuring the radio line 
widths, or by restricting line widths at either the high and/or 
low LW limit, etc.

     The literature on the use of the 21-cm line as a distance 
indicator began with the discovery of the correlation of LW with 
absolute magnitude by \citet{Roberts:69}, \citet{Gouguenheim:69}, and
\citet{Bottinelli:etal:71}. The correlation was then developed as 
a distance indicator by \citet{Tully:Fisher:77}.
   
     The bias properties of individual TF distances for field 
galaxies, chosen from magnitude-limited, or apparent 
diameter-limited samples, is clearly summarized by Teerikorpi in his 
important review. Correction procedures for bias can be made 
either by (1) the method of ``normalized distances'' 
\citep{Teerikorpi:84,Teerikorpi:90,Bottinelli:etal:86,Theureau:etal:97a,Theureau:etal:97b}  
where a ``plateau'' is seen in an $M, (m - M)$ diagram that is 
calculated using TF distances and when all galaxies not on the 
plateau are discarded, or (2) corrected by a method such as the 
``triple entry'' procedure \citep{Sandage:94a,Sandage:94b} or by the
equivalent method of Spaenhauer diagrams 
\citep{Sandage:88a,Sandage:88b,Sandage:94a,Sandage:94b,Sandage:99,FST:94,STF:95}.
 
     The problem of bias for {\em cluster\/} galaxy samples using the
TF method would seem at first to be simpler than for TF field galaxy
samples. Occasional early comments in the literature stated that bias
problems for clusters do not exist because all members of the sample
are nearly at the same distance. This is incorrect. Bias still exists
when a cluster luminosity function is sampled incompletely. 
\citet{Teerikorpi:87,Teerikorpi:90} has called this the 
``cluster population incompleteness bias'', of which there is now a 
significant literature. 

      A listing of the central papers on the incompleteness bias 
include those by 
\citet{Bottinelli:etal:87}, 
\citet{Bottinelli:etal:88a,Bottinelli:etal:88b}, 
\citet*{Kraan-Korteweg:etal:88},
\citet{Fouque:etal:90}, 
\citet{Willick:94}, 
\citeauthor*{FST:94} (their Figs. 5, 6, \& 8), 
\citeauthor*{STF:95} (their Figs. 2, 3, \& 10), 
\citet{Giovanelli:etal:97a,Giovanelli:etal:97b}, 
\citet{Teerikorpi:etal:99}, 
\citet{Masters:etal:06} 
although we disagree with part of their analysis of the bias
\citep*{TSR:07}, and by others cited therein. Why then is
another paper on the cluster incompleteness bias useful, and what is
the justification of this one?

     Clarifications can be made on several points that include 
these. (1). Bias will exist with incomplete cluster sampling due 
to two separate effects, and it is useful to separate them. 

      (a). Applying an incorrect TF slope from the calibrating 
galaxies to the cluster sample is the principal reason for bias. 
Only when the adopted slope for the {\em calibrators\/} is the same as   
that of the {\em sample\/} can a slope bias be eliminated even if the 
cluster luminosity function is sampled ``completely'' in a certain 
way. This is true both for the ``direct'' and the ``inverse'' TF 
formulation, i.e.\ reversing what is the independent and dependent 
variables in the least squares regressions of LW on apparent 
magnitude. 

     The direct and inverse slopes will always be different in 
any correlation of two independent variables if there is 
intrinsic dispersion in the correlation 
\citep[cf.][]{Seares:44,Feigelson:Babu:92}. 

     For the Tully-Fisher correlation it turns out to be crucial 
to use the correct slope in analyzing the data in either the 
direct or the inverse mode. However, there is the question of 
whether we should use the direct slope, the inverse slope, or 
something in between, sometimes called the compromise slope, in 
comparing the observed TF relation with any particular TF 
calibration. 

     (b). The second but related problem is the effect of the 
intrinsic dispersion of the TF correlation itself in the presence 
of incomplete sampling, even if a ``correct'' slope is used. The 
inverse formulation is free from bias in a sampling to different 
apparent magnitude limits 
\citep{Schechter:80,Tully:88,Hendry:Simmons:94,Ekholm:Teerikorpi:97,Teerikorpi:etal:99} 
if the LW distribution is complete (not truncated). However, this 
is true only if the inverse slope is used for {\em both\/} the 
calibrators and the cluster sample rather than the more usual 
practice of using the {\em direct\/} (or the compromise) slope that 
has been determined from the sample but which is then imposed 
upon the calibrators (\S~\ref{sec:04} below) in using the inverse
method for the sample. Rather, the inverse slope must be used for both  
the calibrators and the sample.

     (2). It has sometimes been written that the incompleteness 
bias would disappear if the scatter in the observed TF cluster 
LW-magnitude correlation (i.e.\ either an intrinsic TF dispersion 
or a back-to-front depth effect, or both) would be zero. This 
would only be true if the correct slope for the calibrators is 
used. However, such a slope is never known because of errors in 
the calibrator distances, giving a dispersion in the TF 
calibration solutions. This is the reason for the difference 
between the direct and inverse least squares solutions for the 
calibration. One of the purposes of this paper is to explore the 
effect of this dispersion both in the calibration and in the 
galaxy sample using a variety of TF slopes. 

     The effect of calibrator dispersion on the derived TF slope, 
especially in the direct mode, can be made particularly 
transparent using real calibration data from the HST Cepheid 
database for TF calibrating galaxies. We set out in \S~\ref{sec:04} a
new Tully-Fisher calibration using our HST Cepheid distances 
\citep{Saha:etal:06}. These are the basis of our 2006 distance scale
with its determination of the Hubble constant 
\citep{Sandage:etal:06,TSR:07}. 
   
    (3). A different formulation of the TF bias problems can be 
made and can be compared with earlier discussions that may seem 
to have been unnecessarily opaque \citep[eg.][Paper~II]{Sandage:94b}. 
The present paper aims to clarify the TF bias problem, recasting 
the discussion in Paper~II of this series by retracing the steps 
this author recently took in considering the subject again. The 
purposes here are fivefold. 

     (A). A simplistic toy (zero intrinsic dispersion) model is 
set out in \S~\ref{sec:02} showing what happens to the calculated
distance moduli when an incorrect slope is used for the Tully-Fisher 
correlation for either the calibrators or the cluster data. A 
restriction to avoid a bias is stressed requiring that the depth 
of the magnitude sampling must be faint enough to be symmetrical 
about a ``crossing point'' in the $B_{T}^{0,i}$-LW correlation between
the direct and the inverse regressions if the slope is incorrect. 

     (B). A more complicated toy model with intrinsic dispersion 
is discussed in \S~\ref{sec:03} with and without the correct slopes
for the direct and inverse regressions. 

     (C). Real data are used in \S~\ref{sec:04} to make a mock
(composite) cluster by combining observations for the spirals in the
Virgo~A and B cluster cores with similar data for the Ursa Major I and II 
spiral aggregates, corrected for the small distance differences. 

     (D). \S~\ref{sec:05} sets out a new calibration of the TF relation 
using our HST metallicity-corrected Cepheid distances 
\citep{Saha:etal:06,Sandage:etal:06}. This calibration using various 
slopes is applied to the composite cluster for both the direct 
and inverse regressions in \S~\ref{sec:06}. 

     (E). The resulting distance to the Virgo Cluster A spiral 
core is set out in \S~\ref{sec:07}. The Hubble constant derived from this 
distance, using the cosmic expansion velocity of the cluster, 
freed from all local non-cosmological flows by tying to the 
distant cosmic kinematic frame, is set out there.

\section{MODEL OF THE BIAS CAUSED BY INCORRECT TF SLOPES FOR THE CASE 
         OF ZERO INTRINSIC TF DISPERSION; THE IMPORTANCE OF
         SYMMETRICAL SAMPLING OF THE MAGNITUDE INTERVAL}
\label{sec:02}
\begin{figure}[t]
   \epsscale{0.6}
   \plotone{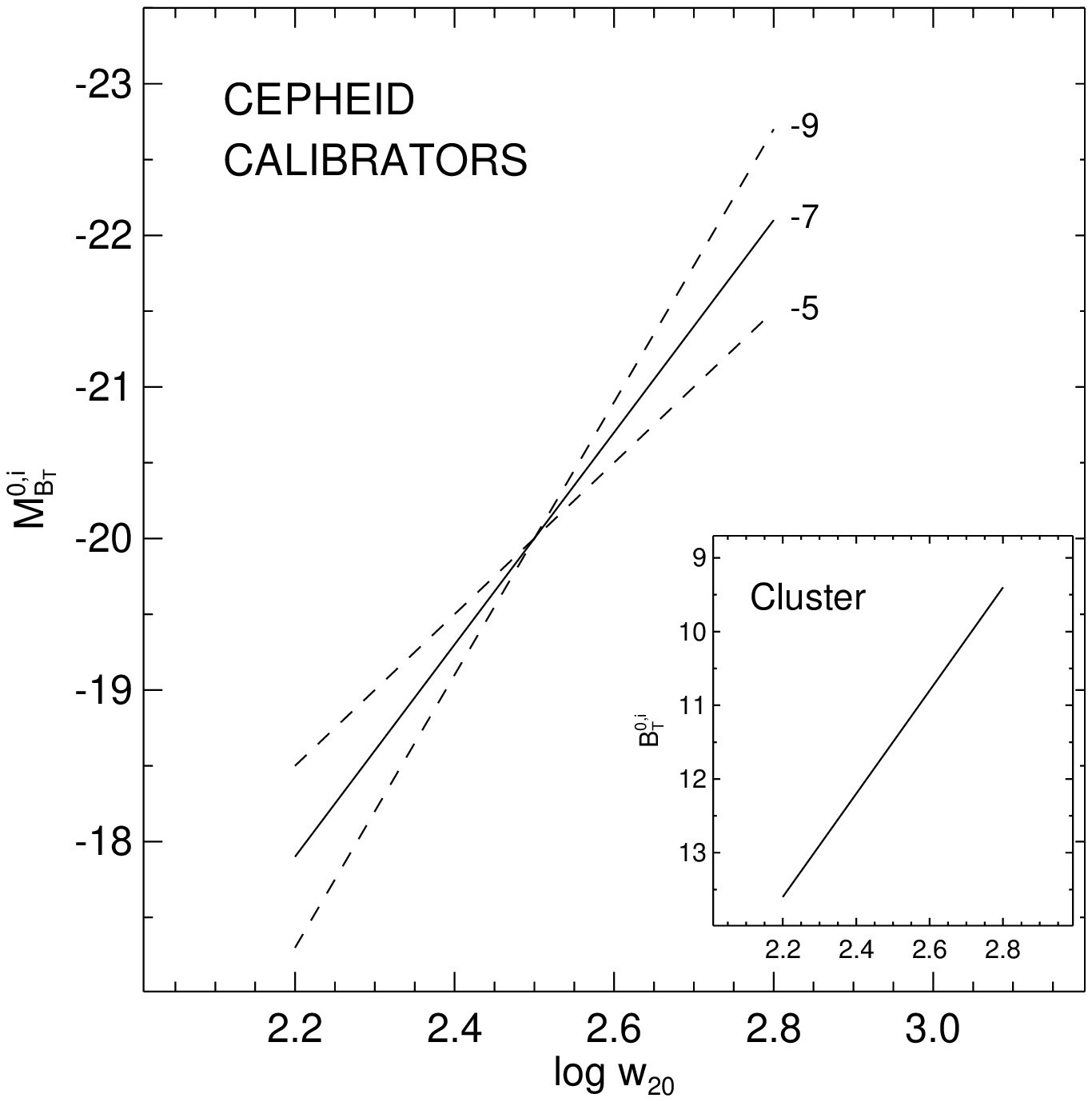}
   \caption{Model of a toy cluster (the insert) with a dispersionless
   TF correlation of line width (corrected for inclination) with
   apparent magnitude that has a slope of 
   $dB_{T}^{0,i}/d\log w_{20}=-7$. 
   A calibration of absolute magnitude with LW in the main body of the
   diagram is shown with variations in the TF slope ranging from 
   $-9$ to $-5$.} 
\label{fig:01}
\end{figure}

As said, even if the intrinsic TF dispersion (and/or an appreciable
depth effect) would be zero, there would yet be a bias in a derived
mean cluster distance if (1) an incorrect slope for the TF correlation
is used, and (2) if the sampling is either incomplete at faint
magnitudes or is non-symmetrical about the cross-over point of the
direct and inverse slopes in the TF diagram. We illustrate these
points using the simplistic model shown in Figure~\ref{fig:01} that
has a zero TF intrinsic dispersion.  


     The insert in Figure~\ref{fig:01} shows a dispersionless TF
correlation for an imaginary cluster that has a true slope of 
$dm/d\log w_{20}=-7$ and that is normalized at $B_{T}^{0,i} = 11.5$ 
for $\log w_{20} = \log W/\sin i = 2.50$. A calibration of the TF
relation with a slope of $-7$ that gives $(m-M)=31.5$ for the toy
cluster is in the main body of the diagram, normalized at absolute
magnitude $-20$ for $\log w_{20} = 2.5$. Two deviant slopes for the
calibration are shown with slopes of $-9$ and $-5$. These cross the
true slope at $M = -20$ and $\log w_{20}=2.5$ by construction. 
These imitate (albeit with exaggeration) the inverse and direct least
squares regressions for real calibration data (Figs.~\ref{fig:05} and
\ref{fig:06} in \S~\ref{sec:04_1}). For real clusters the inverse
correlation is always steeper than the direct. 
(The actual slopes for the 25 galaxy Cepheid calibration in 
\S~\ref{sec:06} are $-6.161$ for the direct regression and $-8.475$
for the inverse, from Table~\ref{tab:07} later).

     We can illustrate the bias on the calculated cluster moduli 
by using each of the calibration (incorrect) slopes of $-9$ and $-5$. 
Consider first the calibration slope of $-9$ applied to the toy 
cluster. This steep slope gives too bright an estimate of the 
absolute magnitude for all line widths larger than the cross-over 
point at $\log w_{20} = 2.5$, and hence the distance moduli are too 
large for all $\log w_{20} > 2.5$ and too small for line widths that 
are smaller than $\log w_{20} = 2.5$. The opposite is true for a slope 
of $-5$; galaxies with $\log w_{20}$ larger than 2.5 will be
incorrectly calibrated too faint, giving smaller (incorrect) distances
for $\log w_{20} > 2.5$ and too bright giving larger (incorrect)
distances for $\log w_{20} < 2.5$.

     In a complete sampling for LWs above and below 
$\log w_{20}=2.5$, and if the number of galaxies along the line in
the insert diagram is constant, then the average distance modulus for
the complete sample will be correct at $(m-M)=31.5$. But this is true 
{\em only if the summing over all apparent magnitudes is carried out
for magnitude limits that are symmetrical about the cross-over
point\/} at $\log w_{20}=2.5$ (or $B_{T}^{0,i}=11.5$ in the toy
model). Said differently, a correct mean modulus will only be obtained
by summing over magnitudes that go to equal limits above and below
11.5, i.e.\ say from 10.0 to 13.0. If the magnitude boundaries are  
not symmetrical about this cross-over point (here at
$B_{T}^{0,i}=11.5$), or if there is a population gradient with
magnitude, then the derived mean modulus will be incorrect. Of course,
also if the summing over magnitude is incomplete (say only over the
restricted interval of 10 to $12\mag$), an incorrect mean modulus will
be derived if the wrong slope is used. 
 
     This is the classical ``population incompleteness'' bias, but 
we see here that it is more complicated than simple incompleteness. It
depends on slope errors and the necessity for symmetrical sampling
above and below the cross-over point of the direct and inverse least
squares solutions even if there is no gradient in galaxies numbers
along the ridge line. This crucial point (see \S~\ref{sec:06}) is
illustrated in Table~\ref{tab:01} for the dispersionless toy cluster
that has been put at a true distance modulus of $(m-M)=31.5$. 


     The first half of the table gives the calculated modulus at 
the fixed magnitudes in column~(1) using arbitrary TF slopes that 
range from $-9$ to $-5$. Fixing all cross-over points of these lines 
to be $M=-20$ at $\log w_{20}=2.5$ gives the equations of the assumed 
lines in the main body of the diagram as $M=-9 \log w_{20} + 2.5$ to 
$M = -5 \log w_{20}-7.5$, etc., and $B_{T}^{0,i}= -7 \log w_{20} +
29.00$.  All entries for $(m-M)$ at the stated apparent magnitudes in
column~(1) of Table~\ref{tab:01} follow in an obvious way from these
equations.

     The calculated moduli for the slopes of $-9$ and $-8$ with 
apparent magnitudes brighter than 11.5 (the cross-over point) are 
larger than the true modulus of 31.5 and visa-versa for fainter 
magnitudes. The opposite trend exists for slopes of $-6$ and $-5$. 
These results are shown in Figure~\ref{fig:02} (top). Of course, for
the ``true'' slope of $-7$, all calculated moduli are 31.5 by 
construction.  


     Using the calculated moduli that are valid {\em at\/} the
specified apparent magnitudes, we can sum the entries to obtain the
average modulus value for all galaxies that are {\em brighter\/} than
any particular limiting magnitude into the cluster luminosity
function, shown in the bottom panel of Figure~\ref{fig:02}. These
moduli are, of course, those that would be obtained in a real case by 
averaging the modulus values for all galaxies in the sample that 
is complete only to brighter magnitude limits. This is the 
incompleteness bias.   

    The effect of an incorrect slope on the bias using incomplete 
sampling is set out explicitly in the second half of
Table~\ref{tab:01}, and shown in the bottom panel of
Figure~\ref{fig:02}. The points to note are these. 

   (A). For steeper slopes than the ``true'' slope (i.e.\ for $-9$ and 
$-8$), the derived mean distance moduli decrease as the sampling is 
done deeper into the cluster luminosity function. The bias effect 
from $B=10$ to 13 is $0.4\mag$ in the modulus differences for the 
slope of $-9$ and $0.2\mag$ for the slope of $-8$. 

   (B). The opposite is true for the shallow slopes of $-6$ and $-5$. 

   (C). The correct modulus is obtained only by sampling to 
$B=13$. This is three magnitudes into the cluster luminosity 
function.

   (D). Sampling fainter than $B=13$ gives incorrect modulus 
values (except for the $-7$ case using the ``true'' slope). This is
the effect of ``non-symmetrical'' sampling about the cross-over 
point.

    Of course, this toy model is too simplistic because there is 
no dispersion in the TF calibration nor in the TF correlation 
for the cluster galaxies. 

\section{A MODEL OF THE BIAS DUE TO AN INTRINSIC TF DISPERSION AND/OR  
         INCORRECT SLOPES FOR THE TF RELATION}
\label{sec:03}
     To understand the bias properties of the cluster TF data it is
useful to analyze the LW-apparent magnitude data in two ways. The
first is to bin the data into intervals of $w_{20}$ and then, within
each interval, to order the listings by apparent magnitude. The second
is to bin the data by apparent magnitude and then to order by line
width within each magnitude interval.  

     The first way is illustrated in Figure~\ref{fig:03} where the 
intrinsic dispersion (plus any back-to-front variation) is shown 
by the dotted envelope lines placed symmetrically about the 
ridge line and where three line-width intervals are drawn for 
illustration. An arbitrary magnitude limit is put at $m = 11.5$, 
which is about $1.5\mag$ fainter than the brightest galaxy in the 
cluster in this example.   


     All galaxies in the line-width interval farthest to the right are
brighter than the magnitude cut-off, whereas galaxies at smaller LWs
to the left are progressively lost to the sample at smaller and
smaller LWs. This causes the bias.        

     Consider first the unbiased LW interval to the right. Due to the
dispersion, some galaxies are brighter than the central ridge line and
some are fainter. If the position of the ridge line has been
calculated by a least squares regression using magnitude residuals at
fixed $w_{20}$, i.e.\ in the ``direct'' solution, there will be equal
numbers of galaxies above and below the calculated ridge line. This
line is the most probable apparent magnitude at that $w_{20}$,
although, because of the dispersion, no galaxy will have the
ridge-line apparent magnitude unless, of course, it is on the ridge
line.

     Consider next how the data in this LW-interval are used to 
obtain a system of distance moduli using some absolute TF 
calibration such in Figure~\ref{fig:08} later in \S~\ref{sec:05}. The
calibration will, itself, have a dispersion that will be a combination
of the intrinsic dispersion due to the physics that governs the TF
correlation and the measuring errors in the distances of the
calibrating galaxies.   

     When the ridge-line calibration (i.e.\ the central line in 
Fig.~\ref{fig:08} later) is applied to the ridge line of the {\em
  observed\/} cluster TF relation such as in Figure~\ref{fig:03}, a
modulus value is obtained of each galaxy in that $w_{20}$
interval. But, as said before, the calculated modulus of any given
galaxy so obtained {\em is incorrect}, unless it is on the ridge line. 

     In detail, consider the unbiased $\log w_{20}$ interval from 2.70  
to 2.75 for the right-hand strip in Figure~\ref{fig:03}. Suppose that
all galaxies in the cluster are at the same distance 
(no back-to-front effect) and there are no errors of measurement
either in $w_{20}$ or magnitude. For this case, galaxies near the
upper envelope line are those that are actually overluminous for their
line width due to an intrinsic dispersion (i.e.\ from the physics) of 
the TF relation. Hence, applying the most probable calibration 
(which is fainter here than these particular galaxies) from the 
ridge-line calibrator relation to such {\em intrinsically\/} bright 
galaxies will give too small a calculated distance modulus for 
them, and visa-versa for galaxies that are actually underluminous 
(i.e.\ below the ridge line). Nevertheless, if the strip is filled 
symmetrically above and below the ridge line, and if the slope of 
that line has been determined by the ``direct'' least squares 
regression, then the mean modulus found by averaging the 
individual moduli in the strip (all of which are incorrect except 
for those on the strip) will be the correct modulus, to within 
statistics.  

     However, if the strip is not sampled completely, as in the 
second and third strips in Figure~\ref{fig:03} where a magnitude cut
truncates the distribution, the average of the individual moduli will
be progressively in error at the smaller line widths as LW intervals 
move toward the left. The average of modulus values for galaxies 
above the magnitude limit line will then be too small by amounts 
that will decrease as the fraction of the accessible sample (the 
ratio to the complete sample) increases as the magnitude limit 
lines are moved fainter. {\em This is the incompleteness bias}. 

     Nevertheless, averaging over all line widths and to a 
sufficiently faint apparent magnitude will give the correct 
distance modulus if, of course, the strips are filled 
symmetrically above and below the most probable ridge line.     

     This bias effect produces an error in the calculated 
{\em slope\/} determined from incomplete data in the direct
regression. The error will be a function of the depth of penetration
into the cluster luminosity function. This is obvious from
Figure~\ref{fig:03}. The slope appropriate for the unbiased LW
interval at the right ($\log w_{20}$ between 2.70 and 2.75) is the
correct direct slope (except for statistics) that would be determined
from the complete sample using the complete luminosity
function. However, if we use only data brighter than any brighter
magnitude cut, here put at $11.5\mag$, the midpoint average magnitudes
for the biased LW intervals  (all to the left of the unbiased interval
in Fig.~\ref{fig:03}) are all brighter than the true ridge line
(unbiased) points shown by the three dots in
Figure~\ref{fig:03}. Hence, the slopes that would be determined  
by calculation using only data brighter than any magnitude cut-off,
{\em will all be too shallow}, approaching the correct slope only 
when the magnitude limit is so faint so as to include the 
complete sample. 

     Hence, because of the incompleteness bias, the calculated 
mean modulus using the direct slope should vary progressively 
with cut-off magnitude. This is the explanation of why in our 
earlier discussion (\citeauthor*{STF:95}, Figs. 2, 3, 6, and 10)
the calculated slope of the observed TF correlation in the direct 
formulation varies with cut-off magnitude. This effect on the 
slope will be demonstrated in \S~\ref{sec:04_1} (Fig.~\ref{fig:07})
using actual data.   

     The ideal schematic model of Figure~\ref{fig:03} is still too
simplistic because the effect of using an incorrect slope for the
ridge line has been ignored. We mean by this that the (direct) slope
shown in Figure~\ref{fig:03}, as it would have been derived from the
complete sample, is assumed to be without error.         

     With this in mind, consider the inverse correlation where 
the calculated TF slope is independent of the depth sampled into 
the cluster luminosity function, in contrast to the direct slope 
calculated from incomplete data. The strategy of analysis for  
the inverse calculation is shown in Figure~\ref{fig:04}. 


     In Figure~\ref{fig:04}, as in Figure~\ref{fig:03}, the scatter of
the cluster data at a given apparent magnitude is shown for a TF
correlation that has intrinsic and/or back-to-front scatter. The
inverse ridge line (marked I) is made to go through the midpoint of
each shaded magnitude interval, because that is what the least squares 
regression gives by its procedure of using residuals in the LWs 
{\em at given magnitudes\/} rather than visa versa. 

     If the distribution of $w_{20}$ within each magnitude interval is
symmetrical about the ridge line, the average distance modulus of 
all the galaxies within each magnitude interval will be the true 
modulus to within statistics. This average at every magnitude 
interval will be bias free. However, as is evident in 
Figure~\ref{fig:04}, this will only be true if the inverse slope is
used. It will {\em not be true if the direct slope is used\/} with the
inverse method.  

     Using either the I or the D ridge lines, and imposing in 
turn their slopes on the calibrator sample, gives individual 
modulus values for each galaxy in any of the individual strips in 
Figure~\ref{fig:04}. Again, as before, these individually are
incorrect, except for galaxies precisely on the ridge
line. Nevertheless, if the distribution of modulus values within each
magnitude interval is symmetrical about the ``I'' ridge line, then the 
average of the individual modulus values within each magnitude 
interval will be the true value provided that the I slope {\em has
been forced on the calibrators themselves}. Clearly, as seen from 
Figure~\ref{fig:04}, the same is not true using the direct
slope. There are more galaxies to the small LW side of the TF
distribution for magnitudes brighter than the cross-over point, and
fewer galaxies to the left of the D line for fainter magnitudes than
the crossover point. This produces the bias in the average modulus using 
the direct slope for all magnitude cut-offs brighter than the 
limit at $B = 14$. This is the cluster incompleteness bias in the 
direct formulation if the sampling is incomplete.   

     In the inverse method we need only average all the modulus 
data in a {\em given magnitude interval\/} to get the correct cluster
modulus, in contrast to the direct method where we need to 
average over the whole of luminosity function, or we must apply 
the correction methods in the direct formulation derived either 
from the ``normalized distance'' method of the French workers 
\citep{Teerikorpi:97}, or by the method of Spaenhauer diagrams 
\citep{Sandage:94a,Sandage:94b}. 

    Figure~\ref{fig:04} also shows the effect caused by an error in
the adopted slope in the inverse method. This will cause a bias in 
the derived distance for the reason just described of using the 
direct slope when the inverse slope should be used.  

     Consider this effect of using the direct slope with the 
inverse method and summing the data only to a given apparent 
magnitude. Consider first all magnitude intervals that are 
brighter than the cross-over point at $m=11.5$ and $\log w_{20}=2.5$. 
An average modulus that is too small will be calculated at each 
of the bright magnitude intervals because there will be more 
galaxies to the left of the adopted ``D'' ridge line than to the 
right (i.e.\ giving moduli values that are too small compared with 
the ridge line average). The opposite will be true for all 
magnitude intervals that are fainter than $m=11.5$ (seen also in 
the top part of Table~\ref{tab:01} for a toy cluster with no
dispersion). Only by summing the total data over all magnitude
intervals can the correct true modulus be obtained. 
However, as in Figure~\ref{fig:01} and Table~\ref{tab:01}, the sum
over all magnitudes in a complete sample must be made no fainter than
the magnitude level that is {\em symmetrical\/} about the cross-over
point, showing again the need to adopt a symmetrical magnitude limit
about the cross-over point in both the direct and inverse solutions.           

     Hence, on both accounts of the effect of (1) incomplete 
sampling using the direct TF slope as in Figure~\ref{fig:03} and then
summing over all magnitudes (usually such completeness is not
available), or (2) of summing over LW to given magnitude limits
(Fig.~\ref{fig:04} for the inverse method), the conclusion is that the
true inverse slope must be used to avoid both (a) the error in the
adopted slope and (b) the incomplete sampling effects. These
conclusions will be demonstrated in \S~\ref{sec:06} using real data
for a composite (artificial) cluster.   

\section{A COMPOSITE CLUSTER MADE TO ILLUSTRATE THE INCOMPLETENESS BIAS}
\label{sec:04}
%
\subsection{Construction of the Composite Catalog}
\label{sec:04_1}
To demonstrate the points made in the last two sections it is useful
to analyze real LW and magnitude data for a galaxy cluster. Only a few
catalogs exist that are complete enough for real clusters to be useful
for such a demonstration. The most complete are for the Virgo~A and B
subclusters and for the Ursa Major ``cluster'', early called the Ursa
Major Cloud by Hubble and identified as such by \citet{Humason:etal:56}
in the Mount Wilson redshift list. 

     However, the number of spirals that are suitable for a TF 
analysis (inclinations greater than $30^{\circ}$ and adequate
photometry to derive fully corrected apparent magnitudes) in these
three aggregates is so small that small-number statistics can mask the
bias effects we are seeking. To increase the size of the sample we
have created a composite cluster by combining the data for the
Virgo~A and B subclusters with those for the Ursa Major cloud to 
form a single catalog. Corrections have been made to the apparent 
magnitudes of the galaxies in the Virgo~B cluster and the Ursa 
Major Cloud to compensate for the small differences in the 
distances of each aggregate relative to Virgo~A.
   
     The data for the Virgo subclusters are taken from the 
listings by \citet{FTS:98} in their Table~3. 
Membership for Virgo~A and B are based on the Virgo Cluster Catalog of
\citet*{Binggeli:etal:85} as reanalyzed for cluster members by
\citet*{Binggeli:etal:93}. Only galaxies with inclination greater than
$30^{\circ}$, and only those assigned either to subclusters A and B
are used. There are 43 galaxies in Virgo~A and 16 in Virgo~B in the
composite catalog.

     The Ursa Major Cloud has been studied in the past by 
\citet{Sersic:60}, 
\citet[][especially his Figs. 8 \& 9]{deVaucouleurs:75} 
in his review of groups, 
\citet{Tully:Fisher:87}, 
\citet{Pierce:Tully:88}, 
\citet[][his Fig. 7 in chapter 4 of his thesis]{Verheijen:97}, 
\citet{Tully:etal:96}, 
\citet{Peletier:Willner:93}, 
\citet{Federspiel:99}, and undoubtedly others. 

     We use here the cluster membership of \citeauthor{Federspiel:99}
from his Table~3.2 for the spirals that are suitable for the TF method 
where he lists the necessary $B_{T}^{0,i}$, line width, and $v_{220}$
radial velocity data (to test membership). Federspiel's membership 
criteria, similar to those of \citeauthor{Verheijen:97} and of
\citeauthor{Pierce:Tully:88}, are (a) heliocentric redshift interval
from 700 to $1500\kms$, and (b) angular distance within $7.5^{\circ}$
of RA(1950)$=11^{\rm h}54^{\rm m}$, Dec(1950)$=+49.5^{\circ}$.
       
     The radial velocity distribution in $v_{220}$, corrected for 
Virgo infall by the model of \citet{Kraan-Korteweg:86}, shows a 
bimodal character with means at $1060\pm29\kms$ and $1485\pm38\kms$,
indicating an appreciable depth effect along the filament connecting
the cloud with the Virgo complex that is shown so remarkably in Figure~6
of \citet{Klypin:etal:03}. We treat separately the two distributions,
assigning galaxies with $v_{220}<1250\kms$ to be members of an Ursa
Major I aggregate, and those with $v_{220}>1250\kms$ as members of
Ursa Major II. There are 18 spirals suitable for the TF method in 
UMa~I and 15 in UMa~II.
     
     Hence, there are 92 galaxies in the artificial composite 
cluster (43, 16, 18, and 15) from Virgo~A, Virgo~B, and UMa~I 
and UMa~II.
         
     The $B_{T}^{0,i}$ magnitudes used here, corrected to a ``total'' 
magnitude, are within a few hundredths of a magnitude of the RC3 
\citep{deVaucouleurs:etal:91} system. (Slight differences at the 
few hundredths mag level exist because no K redshift corrections 
were applied here, differing from the RC3, and some updating was 
made). The LW data for the four aggregates are on the homogeneous 
line-width system used by \citeauthor*{FTS:98}, 
taken from the Lyon-Meudon Extragalactic Data Base (the LEDA). 
These LWs average $0.033\pm0.002\;$dex smaller than the LWs that are  
calculated from the RC3 with the RC3 inclinations calculated from 
the $\log R_{25}$ values listed there. 

     Small magnitude corrections between the four groups are 
needed to reduce all data to the distance of Virgo~A. These were 
determined by comparing the zero points of the least squares TF 
correlations of Virgo~A with those of Virgo~B and UMa~I \& II. To 
calculate these corrections, the Virgo~A TF slope is imposed on 
Virgo~B and UMa~I \& II. These gave zero point magnitude 
differences such that the $B_{T}^{0,i}$  magnitudes for Virgo~B are
made brighter than the \citeauthor*{FTS:98} listings by
$0.38\pm0.16\mag$ (Virgo~B is more distant). The correction for UMa~I
(the smaller velocity aggregate) is $0.16\pm0.13\mag$ made fainter
than the magnitudes listed by \citet{Federspiel:99} (UMa~I is closer
than Virgo~A), and the \citeauthor{Federspiel:99} magnitudes for 
UMa~II (the larger velocity aggregate) were made brighter by 
$0.16\pm0.17\mag$ (UMa~II is more distant than Virgo~A) so as to put
the magnitudes on the distance system as Virgo~A.
      
     Using these magnitude corrections we have made an artificial 
cluster catalogue from which the TF correlation of $\log w_{20}$ with 
the $B_{T}^{0,i}$ magnitudes can be made. The result is in
Figure~\ref{fig:05} where all 92 galaxies of the artificial composite
cluster are shown. A more select subsample of 84 galaxies is shown in
Figure~\ref{fig:06} where a magnitude cut is made at
$B_{T}^{0,i}=14.0$, eliminating the eight faintest galaxies.



     The slopes of the direct and the inverse least squares 
regressions with the cluster data are shown in parentheses in the 
diagrams. The equations of the regression lines are: 
\begin{equation}
B_{T}^{0,i}(\mbox{RC3})=-6.622\log w_{20}\;(\mbox{LEDA})+28.195\pm0.103,
\label{eq:01}
\end{equation}
for the direct regression using all 92 galaxies in Figure~\ref{fig:05}, and 
\begin{equation}
B_{T}^{0,i}(\mbox{RC3})=-8.032\log w_{20}\;(\mbox{LEDA})+31.598\pm0.114
\label{eq:02}
\end{equation}
for the inverse correlation for the same sample. 

     The equations for the more restricted 84 galaxy sample (cut 
at $B_{T}^{0,i}=14.0$) are: 
\begin{equation}
B_{T}^{0,i}(\mbox{RC3})=-5.936\log w_{20}\;(\mbox{LEDA})+26.453,
\label{eq:03}
\end{equation}
for the direct regression, and 
\begin{equation}
B_{T}^{0,i}(\mbox{RC3})=-7.547\log w_{20}\;(\mbox{LEDA})+30.391.  
\label{eq:04}
\end{equation}

\subsection{The Sensitivity of the TF Slope with Sampling Depth into 
             the Cluster Luminosity Function}
\label{sec:04_2}
The data for the artificial cluster have been binned by 
apparent magnitude, enabling a study of the effect on the biased 
slope of the Tully-Fisher correlation due to incomplete 
sampling. Least squares solutions have been made from the 
composite cluster data as the magnitude cut is put progressively 
fainter. The results are listed in Table~\ref{tab:02} in columns (2-6)
for the total sample of 92 galaxies to the limit of $B=15$, and for
the restricted sample of 78 galaxies in columns (7-9) to be discussed 
later in \S~\ref{sec:06}. 


     Column~(1) shows the magnitude cut into the luminosity 
function of the composite cluster. The number of galaxies to each 
truncated limit is in column~(2). The slope and zero point for 
the direct regression (residuals taken in magnitude at a fixed LW 
as in $m = a \log w + b$), are in columns~(3) and (4). The slope and 
zero points of the inverse correlation, as in 
$log w_{20} = cB_{T}^{0,i}+d$, are in columns~(5) and (6). The slopes
and zero points for the complete sample of $n=92$ for $B>15$ (the last
row) are, of course, the same as in equations~(\ref{eq:01}) and
(\ref{eq:02}). Columns~(8) and (9) show the calculated least squares
direct and inverse slopes for the restricted sample ($n=78$) to the
listed cut-off magnitudes. The number of galaxies to that magnitude
limit is in column~(7).  

     Figure~\ref{fig:07} shows the data in Table~\ref{tab:02} where
the progressive increase of slope for the direct correlation,
consistent with Figs. 2, 3, 6, and 10 of \citeauthor*{STF:95},
is evident. The top panel shows the increase of slope with increasing
faintness for the direct formulation using the complete sample of 92
galaxies (col. 3 of Table~\ref{tab:02}). Less clear but consistent
within statistics, is the lack of a correlation of slope with
faintness for the inverse correlation. The bottom panel shows the same
effect more clearly for the restricted sample of 78 galaxies from
columns~(8) and (9) of Table~\ref{tab:02}. 


     The expectation from \S~\ref{sec:03} (eg.\ from
Fig.~\ref{fig:03}) is that there should be a progressive change of
slope as the cluster is sampled more deeply in magnitude using the
DIRECT regression (residuals in magnitude at given LW), whereas no
systematic variation should be present in the INVERSE solution
(residuals in LW at given magnitudes). Figure~\ref{fig:07} shows the
slope effect in the DIRECT (dots) formulation. It is consistent with
Figures 2, 3, 6, and 10 of \citeauthor*{STF:95}. 

\begin{figure}[t]
   \epsscale{0.45}
   \plotone{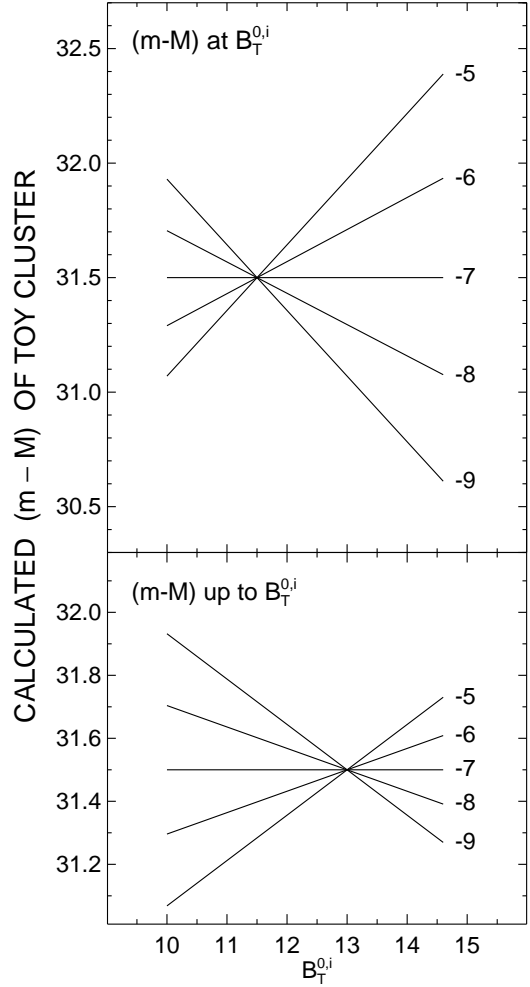}
   \caption{(top): The variation of the derived distance modulus {\em
   at\/} a given sampling magnitude into the cluster luminosity
   function for five assumed TF slope values from data in the first
   half of Table~\ref{tab:01}. All modulus values cross at
   $(m-M)=31.5$ by construction.  
   (bottom): The mean modulus summed to the listed cut-off magnitude
   for the five slope values from the second half of
   Table~\ref{tab:01}. The correct modulus of $(m-M)=31.5$ is obtained
   only when $B_{T}^{0,i}$ stops at 13.0 for all slope values that
   differ from the true slope of $-7$. This magnitude is symmetrical
   about the cross-over magnitude of 11.5 (by construction). Sampling
   fainter than the symmetrical magnitude of 13.0 will give incorrect
   distance moduli, except for the case of the correct slope of $-7$.} 
\label{fig:02}
\end{figure}

\begin{figure}[t]
   \epsscale{0.6}
   \plotone{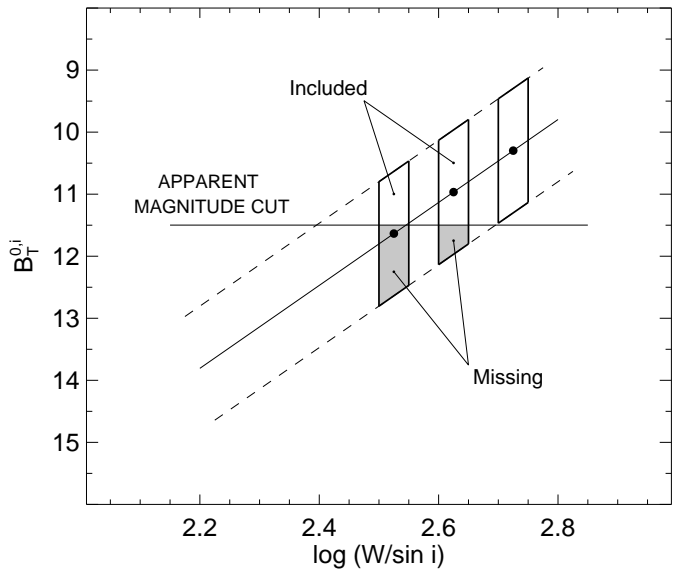}
   \caption{Model of the TF correlation in a cluster with intrinsic
   dispersion and/or an appreciable back-to-front ratio. The bias
   effect of an incomplete sampling of the cluster luminosity function
   is shown, given, for illustration, an observer's magnitude cut-off
   limit at $B_{T}^{0,i} = 11.5$. The right hand vertical LW interval
   is unbiased because none of the data in it are denied entrance into
   the sample fainter; all are brighter than the cut-off
   magnitude. However, the line width intervals to the left are
   progressively biased at this magnitude cut-off. The incompleteness
   bias at any given cut-off magnitude is found by summing over all
   line widths brighter than the cut-off value.}  
\label{fig:03}
\end{figure}

\begin{figure}[t]
   \epsscale{0.6}
   \plotone{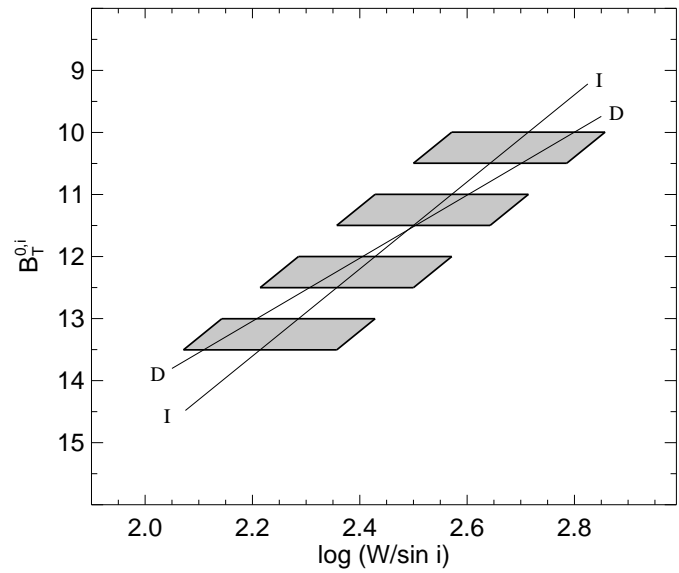}
   \caption{The data for a cluster are binned in apparent magnitude
   intervals. Four magnitude intervals are shown, each enclosed by
   envelope lines that show the intrinsic dispersion 
   (and/or back-to-front effect) in LW at given apparent
   magnitudes. Slopes for both the direct and inverse TF least squares
   ridge lines for the calibrators are shown schematically.} 
\label{fig:04}
\end{figure}

\begin{figure}[t]
   \epsscale{0.6}
   \plotone{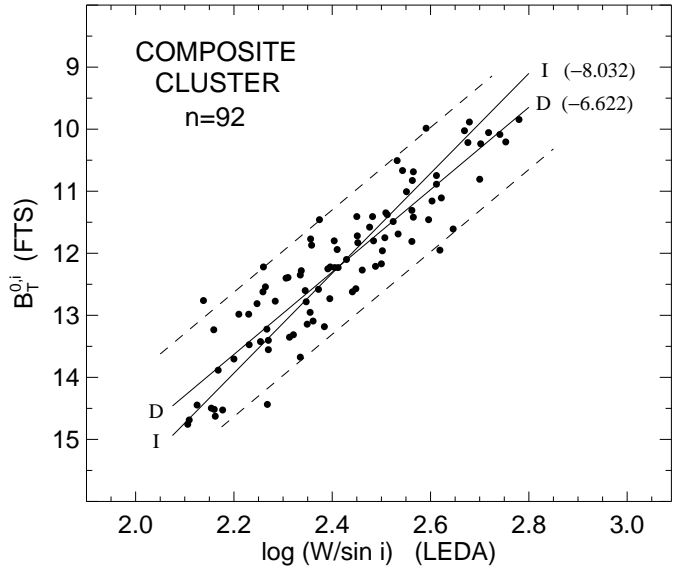}
   \caption{The correlation of line width, corrected for inclination,
   with apparent magnitude for the 92 galaxies in the composite
   cluster composed of galaxies from the Virgo~A and B subclusters and
   the Ursa Major cloud as reduced to the distance of the Virgo~A
   subcluster. The ridge lines for the direct and inverse least
   squares regressions are shown whose slopes are in parentheses.  
   The dashed envelope lines are put one magnitude brighter and
   fainter than the direct ridge line. The magnitudes are on the 
   \citeauthor*{FTS:98} system (their Table~3) which is
   within a few hundredths of a magnitude of the RC3 $B_{T}^{0,i}$
   fully corrected system. The line widths are on the LEDA system
   which is $0.033\pm0.002\;$dex smaller than the RC3 system. The
   equations of the ridge lines are equations~(\ref{eq:01}) and
   (\ref{eq:02}) of the text.} 
\label{fig:05}
\end{figure}

\begin{figure}[t]
   \epsscale{0.6}
   \plotone{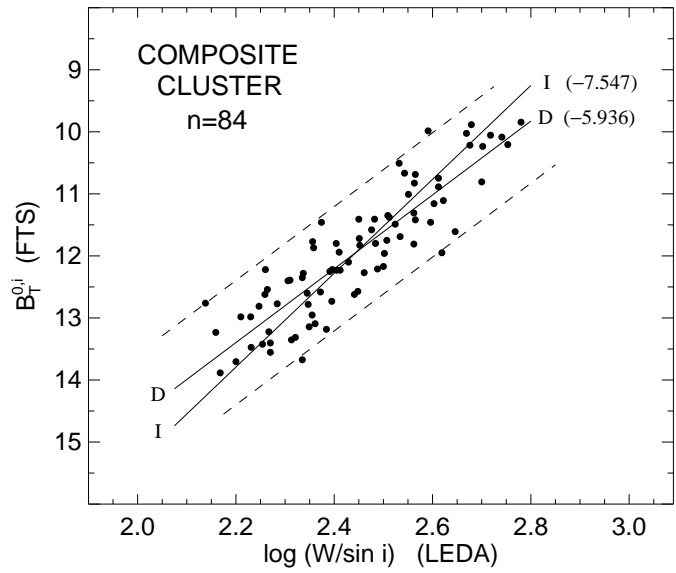}
   \caption{Same as Figure~\ref{fig:05} but for a restricted sample of
   84 galaxies in the artificial cluster, cut from the total sample by
   $B_{T}^{0,i}<14.0$. The equations of the direct and inverse ridge
   lines are equations~(\ref{eq:03}) and (\ref{eq:04}) of the text.}  
\label{fig:06}
\end{figure}

\begin{figure}[t]
   \epsscale{0.6}
   \plotone{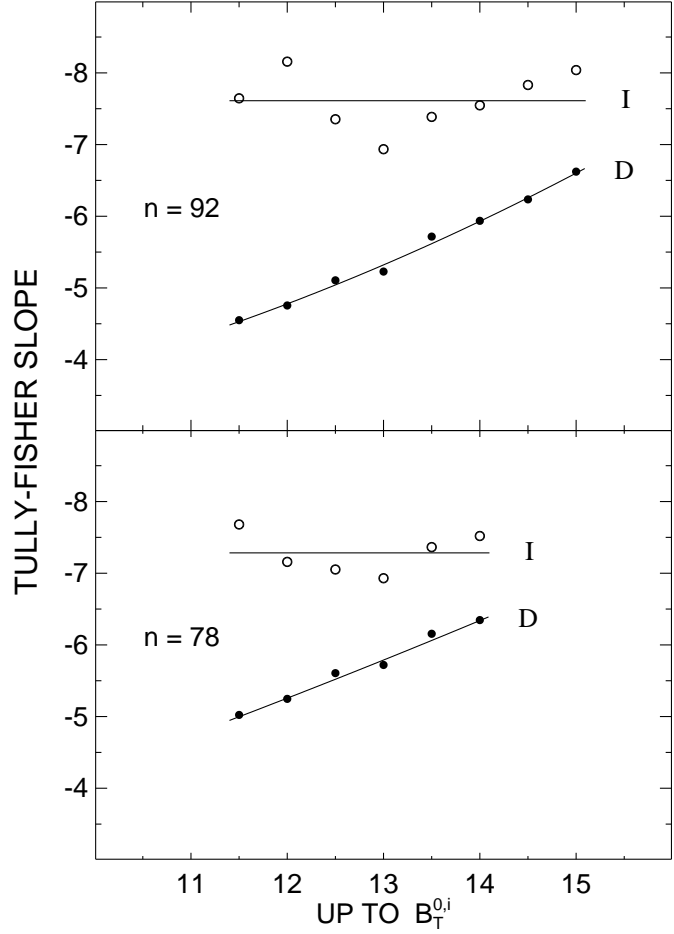}
   \caption{(Top) Change of the Tully-Fisher slope for partial samples
   as depth of penetration into the cluster luminosity function
   increases using data for the total ($n=92$) sample. The DIRECT
   (dots) and INVERSE (open circles) least squares solutions are shown
   as the number of galaxies in the solutions increases from 28 to 92
   and the magnitude cut-offs range from $B_{T}^{0,i}=11.0$ to
   15.0. (Bottom) Same for the restricted sample of 78 galaxies. The
   data are from Table~\ref{tab:02}.  
} 
\label{fig:07}
\end{figure}

\begin{figure}[t]
   \epsscale{0.6}
   \plotone{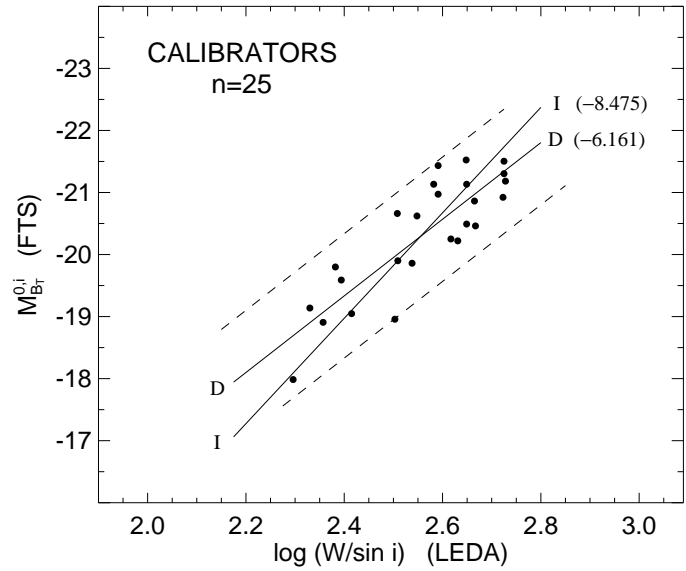}
   \caption{The Cepheid calibrator data from Table~\ref{tab:03} shown
   with the least squares regressions for both the direct and inverse
   formulations whose slopes are in parentheses. These differ from the
   slopes from the data for the composite cluster in
   Figures~\ref{fig:05} and \ref{fig:06}, both of which have higher
   weight than the Cepheid sample alone.} 
\label{fig:08}
\end{figure}

     The situation is less clear in support of the prediction of 
no-slope variation for the inverse regression (open circles) 
because of the dip between 12 and $13\mag$ and the increase at 
fainter magnitudes. However, inspection of the distribution of 
galaxies within the envelope lines in Figures~\ref{fig:05} and
\ref{fig:06} shows the reason is the statistical noise caused by
fluctuations in the small number statistics of the actual data, and
the evident non-uniform filling of the vertical and horizontal bins in
Figures~\ref{fig:03} and \ref{fig:04}, seen by inspection of
Figures~\ref{fig:05} and \ref{fig:06} here, and especially 
Figure~7 of \citeauthor*{FTS:98}. 
In Figures~\ref{fig:05} and \ref{fig:06} for the magnitude interval 
from $B = 12$ to 13 there are more galaxies to the left of the 
ridge lines than to the right, and conversely, from magnitude 13 
to 14 the opposite is true. This causes the entries in the later 
Tables~\ref{tab:05} and \ref{tab:06}, used to construct a
magnitude-modulus diagram similar to Figure~\ref{fig:02}, to show this
statistical noise as curvature from $B = 12$ to 14, reducing the ideal
symmetry of Figure~\ref{fig:02} for the toy cluster to the noisy
reality of the real data for the composite cluster. We shall encounter
the problem again in \S~\ref{sec:06} where the data for the composite
cluster using various TF slopes resembles Figure~\ref{fig:02}. This
statistical noise between $B=12$ to 14 degrades the predictions of the
ideal noiseless toy cluster of Figures~\ref{fig:01} and \ref{fig:02}.      

     To this point we have not needed an absolute magnitude 
calibration of Figures~\ref{fig:05} and \ref{fig:06} to study the bias
problem. In particular we have not needed a calibration to obtain
Table~\ref{tab:02} and Figures~\ref{fig:07}, but to carry the
argument further we now do.

\section{A NEW TF CALIBRATION USING HST CEPHEID DISTANCES FOR 25 GALAXIES}
\label{sec:05}
Equations~(\ref{eq:01}-\ref{eq:04}) and Figures~\ref{fig:05} and
\ref{fig:06} are only useful for determining distance moduli if a
calibration of the TF regression in absolute magnitude is available
from independent data. We use here a new calibration based on 25
galaxies in which Cepheids have been measured, the majority of have
been observed with the HST space telescope. The new data are set out
in Table~\ref{tab:03}.    


    Column~(1) lists the galaxy name. The type and luminosity 
class are in column~(2) from the listings in A Revised Shapley Ames 
Catalog of Bright Galaxies \citep{Sandage:Tammann:87}. 
Column~(3) is the adopted distance modulus from 
\citet[][Table~8]{Saha:etal:06} 
based on period-dependent metallicity corrections required by 
the non-unique Cepheid period-luminosity relations that vary from 
galaxy-to-galaxy 
(\citealt*{TSR:03};
\citealt*{STR:04};
\citealt{Kanbur:Ngeow:04};
\citealt{Ngeow:Kanbur:05};
\citealt{Ngeow:etal:05}). 
These modulus values define our 2006 distance scale based on type Ia
supernovae and HST Cepheids \citep{Saha:etal:06,Sandage:etal:06}. 
(The listed modulus for NGC\,4258 corrects the value listed in
Table~8 of Saha which was a computational error).
       
     Column~(4) is the fully corrected apparent magnitude on the 
scale of \citeauthor*{FTS:98} which is close to that in the
RC3. Column~(5) is the $M_{B(T)}^{0,i}$ absolute magnitude which is
column~(4) minus column~(3). The log line width at the 20\% level and
corrected for inclination is in column (6) on the Lyon-Meudon
Extragalactic Data Base (LEDA) as set out in
\citeauthor*{FTS:98}. Where no data exist in \citeauthor*{FTS:98} we
use the RC3 LW data made smaller by $0.033\;$dex.
  
     The difference in the distance moduli in column~(3) with 
16 galaxies in common with \citeauthor*{FTS:98} is $0.05\pm0.03\mag$ 
with the distances in Table~\ref{tab:03} being larger. 

     Comparisons of the distance scale here with that of 
\citet[][Cols~2 and 5 of their Table~3]{Freedman:etal:01} are these. 
The difference between Table~\ref{tab:03} here and Table~3, column~(2)
of \citeauthor{Freedman:etal:01} (their old scale) is
$0.09\pm0.04\mag$ from 25 overlaps. Our distances are larger. The
difference using the new scale of \citeauthor{Freedman:etal:01} (their
Table~3, col. 5) is $0.23\pm0.04\mag$, again our distances are larger.
 
     Comparing the larger sample of all Cepheids (not just those 
that make up the TF calibration) that are in common between our 
scale and Freedman shows $(m - M)=0.11\pm0.03\mag$ for 
Freedman's old scale (col.~2 of their Table~3) from 30 overlaps. 
Their new scale (col.~5 of their Table~3) compared with ours  
shows $(m-M)=0.25\pm0.04\mag$ from 30 galaxies in common. 
Again our scale is longer.

     The reason for the difference is that we account for the 
different slopes and zero points of the Cepheid P-L relations 
from galaxy-to-galaxy according to metallicity corrections that 
are period dependent \citep{Saha:etal:06}. The P-L slopes differ 
depending on metallicity 
\citep[Fig.~5 and Table~4 of][]{TSR:07}, whereas those of
\citeauthor{Freedman:etal:01} do not. For their new scale 
(column~5 of their Table~3), they adopt the shallow LMC P-L 
slope, not the steeper slope for the Galactic Cepheids which we 
have argued elsewhere \citep{TSR:07} that they should have 
used. 

    The data from Table~\ref{tab:03} are plotted in
Figure~\ref{fig:08} where the direct and inverse least squares
regressions are shown with their slopes in parentheses. These slopes
differ from the regressions for the composite cluster in
Figures~\ref{fig:05} and \ref{fig:06}. They {\em must not\/} be used
in analyzing the cluster data. Rather, the cluster slope must be 
imposed on the calibrator data (\S~\ref{sec:06}) to obtain the
calibrator zero point that is appropriate for the cluster data. 
This requirement is fundamental to avoid bias errors due to slope 
differences. This crucial point concerning the correct slope to 
use has sometimes been overlooked in the earlier literature on 
the TF method.   


\section{BIAS IN THE DERIVED DISTANCE MODULUS DUE TO INCOMPLETE 
    SAMPLING IN THE ARTIFICIAL CLUSTER USING BOTH THE DIRECT 
        AND THE INVERSE TULLY-FISHER REGRESSIONS }
\label{sec:06}
Individual distance moduli have been calculated for all 
galaxies ($n=92$) in the artificial cluster using both the direct 
and inverse slopes. The calibration using the Cepheid data in 
Table~\ref{tab:03} has been zero pointed by imposing the cluster slopes of    
$-6.622$ and $-8.032$ from equations~(\ref{eq:01}) and (\ref{eq:02})
on the Cepheid calibrators. These give the Cepheid ridge-line
calibrations of    
\begin{equation}
M_{B(T)}^{0,i}(\mbox{FTS}) = -6.622 \log w_{20}\;(\mbox{LEDA})-3.369\pm0.103
\label{eq:05}
\end{equation}
for the direct regression, and 
\begin{equation}
M_{B(T)}^{0,i}(\mbox{FTS}) = -8.032 \log w_{20}\;(\mbox{LEDA})+0.239\pm0.114
\label{eq:06}
\end{equation}
for the inverse. 

     The individual (most probable, but all incorrect because of 
the dispersion except for galaxies on the ridge lines) moduli 
follow in an obvious way by combining these Cepheid ridge-line 
calibrations with the individual apparent magnitude and LW data 
for each galaxy in the cluster.

     Following Figure~\ref{fig:03}, the data were first separated into
$\log w_{20}$ intervals in steps of $0.1\;$dex. They were then binned
in apparent magnitude intervals of $0.5\mag$ and averaged. 

     The average modulus in each $w_{20}$ interval and each magnitude 
bin are shown in Table~\ref{tab:04}. The number of galaxies that make up 
each average is in parentheses.                                         


     The bias properties discussed in \S~\ref{sec:03} can now be 
illustrated using Table~\ref{tab:04} rather than by relying on
hypothetical models. These properties and their consequences can most
easily be understood by using Table~\ref{tab:04} and referring to
Figures~\ref{fig:03} and \ref{fig:04} for the two binning
strategies. There are three principal points.    

     (1). Table~\ref{tab:04} shows that within any given LW interval 
(columns 2-8) the average moduli show a systematic variation as the
apparent cut-off magnitude is made fainter. This is expected from the
discussion of Figure~\ref{fig:03} in \S~\ref{sec:03}. The variation is
due to this.
   
\begin{figure}[t]
   \epsscale{0.75}
   \plotone{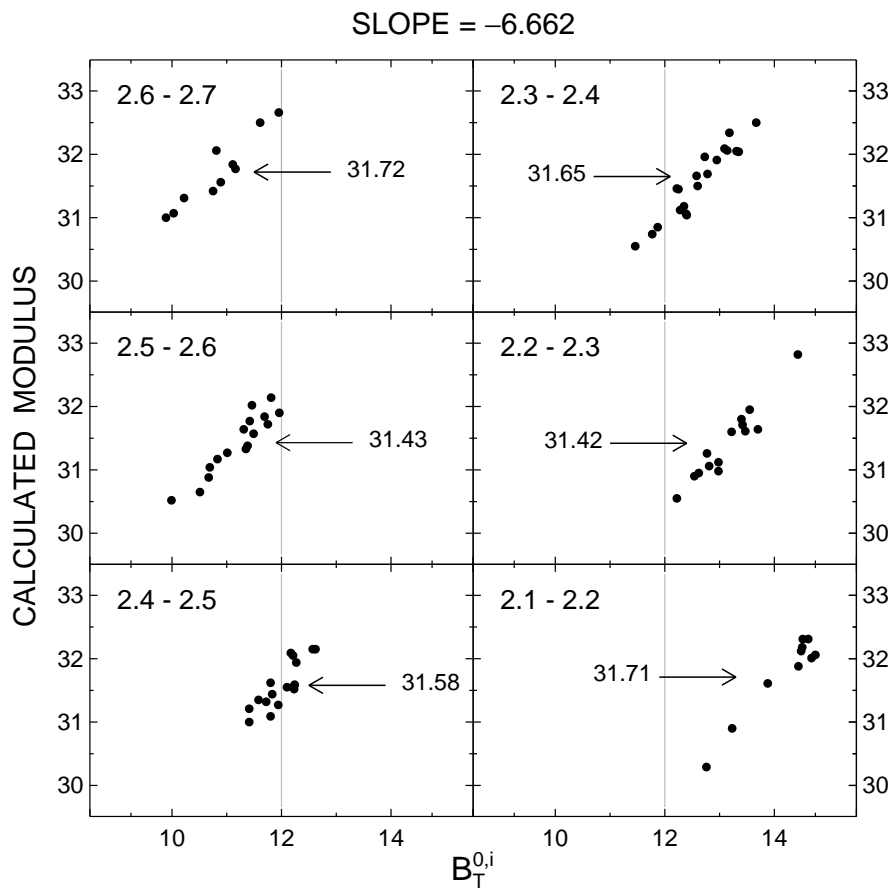}
   \caption{The variation with apparent magnitude of the calculated
   distance moduli for each galaxy in the artificial cluster when
   binned in $\log w_{20}$ intervals in steps of $0.1\;$dex using the
   DIRECT slope of $-6.622$ in Table~\ref{tab:04}A for the complete
   sample of $n=92$. The mean modulus marked by the arrow in each
   panel is that listed at the bottom of Table~\ref{tab:04}A. A light
   vertical line to guide the eye is put in each panel at $B=12.0$.} 
\label{fig:09a}
\end{figure}

     All galaxies in Figure~\ref{fig:03} that are brighter than the
ridge line magnitude at any LW have calculated moduli that are
{\em smaller\/} than the ridge-line (the true) value and visa versa
for those below the ridge line. Figures~\ref{fig:09a}a and
\ref{fig:09b}b show the trend of the individual moduli for both the
direct and inverse slopes in the six LW intervals that range from 
2.6-2.7 to 2.1-2.2.   


     The expectation (\S~\ref{sec:03}) that becomes manifest in
Table~\ref{tab:04}A  is that the average over all magnitudes
(i.e.\ summed vertically in each column of Table~\ref{tab:04}A) should
not vary systematically {\em with LW\/} to within statistics using the
DIRECT slope because the DIRECT ridge line threads strictly down the
middle of each vertical column of Figure~\ref{fig:03} by its least
squares construction. This expectation is realized to within
statistics by the averages set out at the bottom of
Table~\ref{tab:04}A for each LW interval.   


     In contrast, using the INVERSE slope, the averages over all 
magnitudes {\em at given LW}, as analyzed via Figure~\ref{fig:04}, is
expected to vary systematically with LW, meaning that the averages at
the bottom of Table~\ref{tab:04}B should show a strong systematic
variation with LW because the INVERSE ridge line, not shown in
Figure~\ref{fig:03}, {\em does not\/} thread the middle of the
vertical LW columns. This expectation is well seen in the strong
variation, outside the statistical noise, of the average moduli shown
in Figure~\ref{fig:09b}b marked by the arrows, and  
seen in the bottom row of Table~\ref{tab:04}B.   

     (2). However, rather than first binning by LW and then 
magnitude, the more realistic analysis, and the one always made 
in practice, is to work with averages made by summing over all 
line widths at various magnitude limits. The strategy is  
illustrated in Figure~\ref{fig:04} using progressively fainter magnitude 
cut-offs. Here, we are interested in the bias properties of 
Tables~\ref{tab:04}A and B by summing over 
{\em all line widths\/} to various cut-off magnitude limits. The method
using Tables~\ref{tab:04}A and B is to sum  
horizontally (over all LW) in each magnitude interval, and than 
sum over all magnitude intervals up to a given magnitude cut-off. 
 
    The results are given in Table~\ref{tab:05} which is divided using
the direct slope of $-6.622$ in the top half and the inverse slope of   
$-8.032$ in the bottom half. Column~(1) shows the magnitude intervals 
used in Table~\ref{tab:04}. Column~(2) is the number of galaxies used
in the averages. Column~(3) gives the average modulus for all galaxies 
summed over all LWs in this magnitude interval. These are calculated
by summing the Table~\ref{tab:04} entries horizontally in each  
magnitude interval. Column~(4) is the rms variation of these
averages. Columns (5-8) give the average moduli calculated by summing
the data in column~(2) {\em up to\/} the magnitude cut-off listed  
in column~(5). The number of galaxies making up this sum is in 
column~(6). This is the accumulated number of galaxies by summing 
column~(2). The mean modulus of the accumulated sums is in 
column~(7). The rms of these averages is in column~(8).  


   (3). The expectation from \S~\ref{sec:03} is that the inverse slope
must be used in summing horizontally in Table~\ref{tab:04} as in
Figure~\ref{fig:04} to avoid bias. If so, there should be no
systematic bias increase in the derived distance modulus based on the
inverse slope as the grasp into the cluster luminosity function is
deepened. However, this should not be true in the direct formulation
where the derived average modulus is expected to increase with fainter
magnitude cut-offs. 

     These expectations are verified in Figure~\ref{fig:10} where the
data from Table~\ref{tab:05} are plotted. Open circles are for the
magnitude intervals in column~(1). Crosses are for the accumulated
data listed in column~(7). The expectations from the discussion in 
\S~\ref{sec:03} are fulfilled to within statistics in the two panels
of Figure~\ref{fig:10}.   


\subsection{Effect of Using an Incorrect Slope for the TF Correlation}
\label{sec:06_1}
Table~\ref{tab:05} and Figure~\ref{fig:10} are based on the formal
least squares slopes for the TF data for the composite cluster for
both the direct and inverse calculations. It is of interest to gauge 
the effect of systematic errors in these slopes by arbitrarily 
varying them over a wider range than given by the regressions on 
the actual. 

     We have chosen TF slopes of $-9$, $-7$, $-6$, and $-5$ to add to
the results in Table~\ref{tab:05}. These cover the same range of
slopes as in the toy cluster of Figures~\ref{fig:01} and
\ref{fig:02}. The calculation was made by first redoing the Cepheid
zero point calibration by imposing each of the assumed slopes on the
Cepheid data in Table~\ref{tab:03}. These $M$, $\log w_{20}$ 
zero point calibrations were then applied to each of the 92 
galaxies giving individual $(m-M)$ modulus values for each cluster 
galaxy. These were then binned by apparent magnitude and summed 
over all LW for the galaxies brighter than a given apparent 
magnitude to produce a mean modulus for that TF slope and 
apparent magnitude cut-off. 
The results are listed in Table~\ref{tab:06}, which can be combined
with those in Table~\ref{tab:05} for the slopes of     
$-6.622$ and $-8.032$ and Table~\ref{tab:08} later for slopes of 
$-6.346$ and $-7.519$.


     Column~(1) is the magnitude depth that is sampled into the 
cluster luminosity function. The number of galaxies making up the 
averages is in column~(2). The average moduli to the cut-off 
magnitude are in columns (3), (5), (7), and (9). The rms of these
listed moduli are in columns (4), (6), (8), and (10).   

     Table~\ref{tab:06} for the slopes of $-9$ to $-5$ combined with
Table~\ref{tab:05} for the data with the slopes of $-6.622$ and
$-8.032$ show many of the general features of Figure~\ref{fig:02} and
Table~\ref{tab:01}.  

     As expected, the systematic run of moduli with apparent 
magnitude exists for all slopes that differ from the inverse 
slope of $-8.032$. For slopes steeper than this ($-9$ in
Table~\ref{tab:06}) the sign of the bias is that smaller moduli are
calculated as the magnitude grasp into the cluster is increased. 
The opposite sign of the bias is evident (larger moduli for deeper
cluster penetration) is seen for slopes flatter than $-8.032$. 
 
    However, the detailed results, seen by plotting the data in 
Tables~\ref{tab:05} and \ref{tab:06} (not shown), are not as clear as 
in the noiseless model of Table~\ref{tab:01} and
Figure~\ref{fig:02}. From Figure~\ref{fig:02} (bottom) we expect that  
the null magnitude (where all modulus values converge) should be 
at $B=13$, which is about $3\mag$ fainter than the brightest 
galaxy. However, the results for the composite cluster using real 
data show that the null magnitude is near $B=12$ from
Table~\ref{tab:06} and between $B=12$ and 12.5 from
Table~\ref{tab:05}. As explained in \S~\ref{sec:04_2},  
the small-number statistical noise in the distribution of points 
between $B=12$ and 14 is the probable cause of this difference. 
This shows the limitation on the accuracy of the TF method due to 
random noisiness within the dispersion when only a small number 
of cluster galaxies are available. Of course, the situation is 
improved using $I$ magnitudes where the dispersion is smaller but 
the slope of the TF regression is near 10, adding additional 
uncertainties. 

\section{DISTANCE TO THE VIRGO A SPIRAL CLUSTER CORE LEADING TO 
           THE GLOBAL VALUE OF THE HUBBLE CONSTANT}
\label{sec:07}
From the results concerning bias set out in the previous 
sections, how then do we use them to find the most probable 
distance modulus of the composite cluster? 

     From analysis of the properties of the bias given above and 
from the results of Figures~\ref{fig:07}, \ref{fig:10}, and
\ref{fig:11} it might seem best to use only the moduli based on the
inverse analysis because, given large enough samples, there is no
bias. However, there are two reasons we have not done this. 
 
    (1). The rms variation of the modulus summed to various 
magnitude limits, listed in Table~\ref{tab:05} and later in
Tables~\ref{tab:07} and \ref{tab:08} in \S~\ref{sec:07_1}, is always
smaller for the direct formulation than for the inverse in a ratio of
about 1.2 (eg.\ 0.576 to 0.488 from the penultimate rows of
Table~\ref{tab:05}). Hence, the direct moduli have greater weight than
the inverse by a factor of about 1.4.  

     (2). There are always differences in the final average moduli 
between the direct and inverse calculations that are independent 
of the bias problems. The cause is different Cepheid calibrations 
between the direct or inverse calculations. This is because a 
given galaxy will be assigned different absolute magnitudes 
depending on the two different calibrations of zero point. When 
the sums over all galaxies are taken, these will not average out
because the direct and inverse magnitudes have different weights. 
The experience here (Table~\ref{tab:05} summed to $B=14$,
Table~\ref{tab:07} col.~7 in \S~\ref{sec:07_1}, and
Table~\ref{tab:08} also later in \S~\ref{sec:07_1}), the mean modulus 
taken over all cluster galaxies is always larger for the direct 
calculation than for the inverse. The modulus difference between 
the direct and inverse is always smaller when their slope 
difference is smaller, seen from the second half of
Table~\ref{tab:07}.  

     In view of these two reasons, which regression should we 
use to obtain the best distance modulus? We have decided not to 
decide between them but to reduce the data with both, avoiding 
the bias corrections for the direct method by summing to $B=14$, 
a magnitude that is symmetrical brighter and fainter than the 
cross-over point of the direct and inverse regressions.     

\subsection{The Cepheid and Cluster Zero Point Calibrations Using 
                  Various TF Slopes }
\label{sec:07_1}
For each adopted TF slope for the cluster sample we must 
determine new zero points for the Cepheid calibration based on 
Table~\ref{tab:02} of \S\,\ref{sec:02} by forcing the adopted slope on
the calibrators and calculating the resulting intercept for the
calibration equation. (As said before, this point is sometimes missed
in the literature where a same slope is often used both for the direct 
and inverse method in the Cepheid calibration even as these 
slopes differ from the slopes that apply to the cluster data).  

     In the preceding sections we have used a variety of slopes, 
some arbitrary and others calculated from the Cepheid data, and 
some from the composite cluster with various magnitude
truncations. Table~\ref{tab:07} summarizes all of the choices
discussed in the previous sections.  


     The equations of the regressions in Table~\ref{tab:07} are 
$M = a \log w_{20} + b$ and $B = a \log w_{20} + c$ where the $a$,
$b$, and $c$ values are listed in columns (1), (2), and (4) of the
table. Column~(3) is the rms of the absolute magnitude zero point
calibration, $b$, in column~(2).  
Column (5) is the rms of the apparent magnitude zero point in 
column~(4). Column (6) identifies the sample used, ranging from 
$n=92$ for the total sample, $n=84$ for the total sample truncated 
at $B=14.0$, to $n=78$ for a sample still further truncated by 
eliminating the four outliers on the bight side and the two 
outliers on the faint side near the limit lines in
Figures~\ref{fig:05} and ~\ref{fig:06}.  
Column~(7) shows the mean distance modulus using the cluster data 
with the listed $b$ and $c$ calibrations and the magnitude limits for 
the cluster data in column~(8). The moduli in column~(7) are the 
differences between columns (4) and (2). The first five rows are the 
results for the run of assumed slopes discussed in \S~\ref{sec:06_1}. 

     The strong dependence of the final moduli (summed to the 
cluster magnitude limit over the range of $B$ from about 10 to 
$15\mag$) on the slope of the TF regression is shown by the variation 
in column (7) for the slopes in first five rows of the table. For 
the more restricted range of the slopes shown by the four pairs 
of values in the last part of the table, the modulus values range 
from 31.30 to 31.60. The values for the highest weight sample 
(i.e.\ that with the lowest rms values) with $n=78$, vary from      
$(m-M) = 31.42$ for the inverse solution to 31.56 for the direct 
solution.
  
     It is of interest to set out the data for this highest 
weight sample in the same detail as in Table~\ref{tab:05} for the
complete sample. Table~\ref{tab:08} lists the partial moduli in the
various magnitude intervals and magnitude limits for the restricted
sample format for the direct slope of $-6.346$ and the inverse slope
of $-7.519$. Table~\ref{tab:08}A shows the results in magnitude
intervals; Table~\ref{tab:08}B shows the results of summing to various
magnitude limits.


    The data are plotted in Figure~\ref{fig:11}. The open circles are
from Table~\ref{tab:08}A; the crosses from Table~\ref{tab:08}B. 
The noise in the correlations is caused by the nonuniform filling of
the correlation between $B=12$ and 14, mentioned earlier.  


\subsection{The Adopted Distance Modulus for the Virgo~A, Virgo~B and 
               Ursa Major I and II Clusters }
\label{sec:07_2}
From Table~\ref{tab:08} and Figure~\ref{fig:11} we adopt as our final
TF modulus for the composite cluster the mean of the direct and inverse 
calculation using the $n=78$ restricted sample. The values from 
Table~\ref{tab:08}B are $(m-M)=31.56\pm0.05$ for the direct calculation 
and $(m-M)=31.41\pm0.05$ for the inverse, giving 
$\langle m-M\rangle=31.49\pm0.04$ as the weighted mean. 

     Following \citeauthor*{FTS:98} (\S~8 there) we apply a systematic
correction of $0.07\mag$ (reducing the modulus) to account for the
fact that the cluster members at a given LW are redder on average  
(and also are hydrogen deficient) than for the calibrators. 
Hence, our final TF modulus of the Virgo~A cluster is 
$(m-M)=31.42\pm0.04$ (internal). 

      Applying the magnitude offset from Virgo~A to Virgo B of
$0.38\pm0.16\mag$ (Virgo~B is more distant than Virgo~A), of 
$0.16\pm0.13\mag$ for UMa~I made closer, and 
$0.16\pm0.17\mag$ for UMa~II made further from \S~\ref{sec:04_1} gives
the modulus values of these clusters as 
$(m-M)=31.80\pm0.16$ for Virgo~B, 
$31.26\pm0.13$ for UMa~I, and 
$31.58\pm0.17$ for UMa II.  
   
\subsection{Systematic Errors}
\label{sec:07_3}
The formal (internal) error of $\pm0.04$ for Virgo~A is, of course, 
too small because of systematic errors in 
(1) the Cepheid P-L relation of the order of $0.1\mag$ in its 
zero point \citep{TSR:07}, and the error in the zero point
of equations~(\ref{eq:05}) and (\ref{eq:06}) here, 
(2) the uncertainty of $\sim\!0.1\mag$ in the zero point of the 
Cepheid TF regression in Figure~\ref{fig:08} (cf. the Cepheid rms
value in Table~\ref{tab:07}), and (3) the statistical noise due to the
nonuniform filling of the TF correlation in Figures~\ref{fig:05} and
\ref{fig:06} as discussed above. It is impossible to put an exact
value on the systematic  accuracy of $(m-M)$ due to these causes, but
a reasonable estimate is that it is no smaller or larger than
$\pm0.2\mag$. Hence, we adopt $(m-M) = 31.42\pm0.2\mag$ as our final
modulus of Virgo~A using the Tully-Fisher method.  

     If, as is sometimes done, we combine Virgo~A and B, treating  
them as a single cluster with $(m-M) = 31.42\pm0.2$ (assumed) from 
43 galaxies in Virgo~A and $(m-M)=31.80\pm0.2$ (assumed) from 
Virgo~B from 16 galaxies, the weighted mean of A and B is 
$(m-M)=31.52\pm0.14$ (external assumed), which, however, we do not
use. For comparison, the value derived by \citeauthor*{FTS:98} (their
eq.~11) where clusters A and B are combined, is $31.58\pm0.24$
(external).  
 
\subsection{The Hubble Constant from the Adopted Virgo~A Distance}
\label{sec:07_4}
From the discussion by \citeauthor*{FTS:98} (Fig.~14 and \S~11.2), the 
expansion velocity of the Virgo core, freed from all streaming 
motions and tied to the external cosmic frame beyond the local 
bubble whose edge is near $4000\kms$ 
(\citealt{Jerjen:Tammann:93}; \citeauthor*{FST:94}, Figs. 17-19;
\citealt{Masters:etal:06}), is $1175\kms$. This is derived from a high
weight solution of the relative Hubble diagram (velocity vs. distance
{\em ratios}) by \citet{Jerjen:Tammann:93}
where the Virgo core is tied to 17 more remote clusters augmented 
by 24 clusters from Giovanelli (private communication in 1996) 
who also used distance ratios to Virgo.            

     Taking Virgo~A to be the core relative to which the 
distance ratios to the 41 galaxies pertains, and using 
$(m-M)=31.42\pm0.2$ ($D=19.2\;$Mpc, with a range from 17.5 to $21.1\;$Mpc) 
combined with $1175\kms$ for the free expansion velocity, the
resulting Hubble constant is $H_{0}=61\ksm$ with a range from  
53 to 70, where we have also put an error of $\pm50\kms$ on the 
Virgo~A expansion velocity. 

     If, for some reason, one does not wish to rely on the 
distance ratios to Virgo of \citet{Jerjen:Tammann:93}, as 
augmented by Giovanelli's private communication in 1996, another 
route to obtain the free expansion velocity of the Virgo core is 
via a velocity perturbation model for the Virgo velocity relative 
to the remote frame. The observed mean heliocentric velocity of 
the Virgo core is $1050\pm35\kms$ \citep{Binggeli:etal:93}. This
transforms to $932\kms$ relative to the centroid of the Local Group
\citep*{Yahil:etal:77}. Using $v(\mbox{infall})=220\kms$
\citep{Kraan-Korteweg:86} for the Virgo infall velocity 
gives $v_{\rm Virgo}=1152\pm35\kms$ which is the same as $1175\kms$ 
to within statistics.     

     The value of $H_{0}=61$ obtained here agrees well with the 
value determined via the supernovae route in 25 separate studies, 
most of which give $H_{0}$ between 50 and 70 
\citep[summarized in Table~6 of][]{Sandage:etal:06} 
by a plethora of authors. However, we consider the method here to be
of considerably lower weight than that using supernovae because of the
evident problems raised by the population incompleteness bias when
using clusters and the problem of tying the Virgo~A cluster accurately
into the remote cosmic kinematic frame.     



\clearpage
\clearpage
\clearpage
\clearpage
\clearpage
\clearpage
\clearpage
\clearpage
\clearpage
\makeatletter
\def\fnum@figure{{\rmfamily Fig.\space\thefigure a.---}}%
\makeatother
\clearpage
\setcounter{figure}{8} 
\makeatletter
\def\fnum@figure{{\rmfamily Fig.\space\thefigure b.---}}%
\makeatother
\begin{figure}[t]
   \plotone{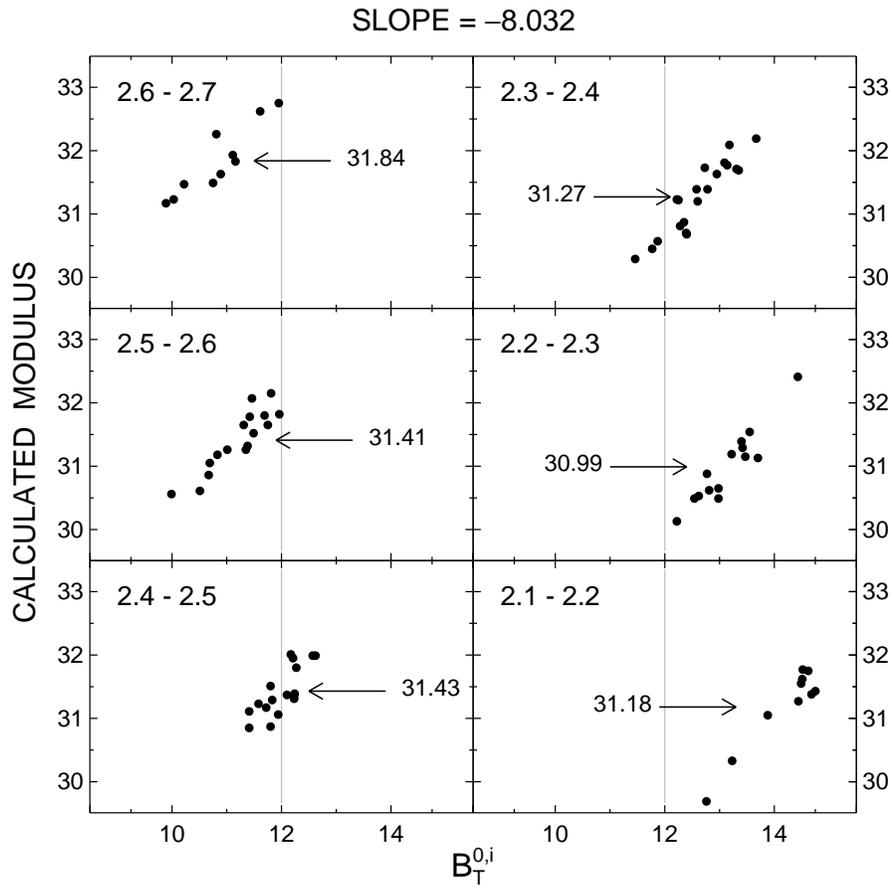}
   \caption{Same as Figure~\ref{fig:09a}a using the INVERSE slope of
   $-8.032$ and the individual data for the composite cluster.} 
\label{fig:09b}
\end{figure}
\makeatletter
\def\fnum@figure{{\rmfamily Fig.\space\thefigure.---}}%
\makeatother
\clearpage
\begin{figure}[t]
   \epsscale{0.6}
   \plotone{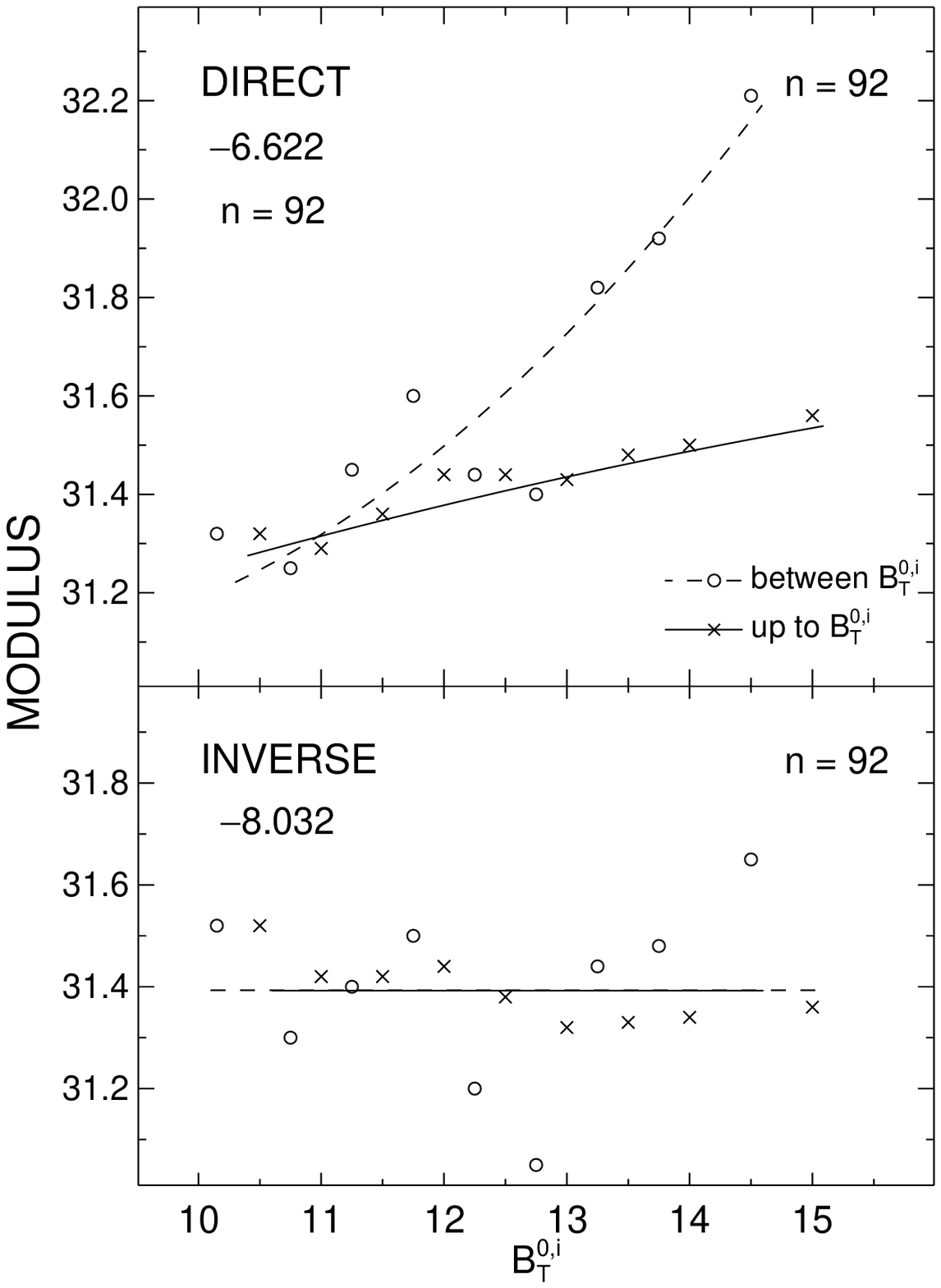}
   \caption{(Top). The average moduli for the total sample ($n=92$)
   listed in Table~\ref{tab:05} in each magnitude interval using the
   direct slope of $-6.622$ and summed over all line widths {\em in\/}
   that magnitude interval (open circles), and then summed over 
   {\em all\/} magnitude intervals (crosses) up to the listed 
   $B_{T}^{0,i}$ magnitude cut-off limit (crosses). (Bottom) Same for
   the inverse  formulation with a slope of $-8.032$ for the total
   sample. The noise in both the open circle and the crosses data
   between $B=12$ and 13 is caused by nonuniform filling of the TF
   distribution between the envelope lines of Figures~\ref{fig:05} and
   \ref{fig:06}. The data are from Table~\ref{tab:05}.} 
\label{fig:10}
\end{figure}
\clearpage
\begin{figure}[t]
   \epsscale{0.6}
   \plotone{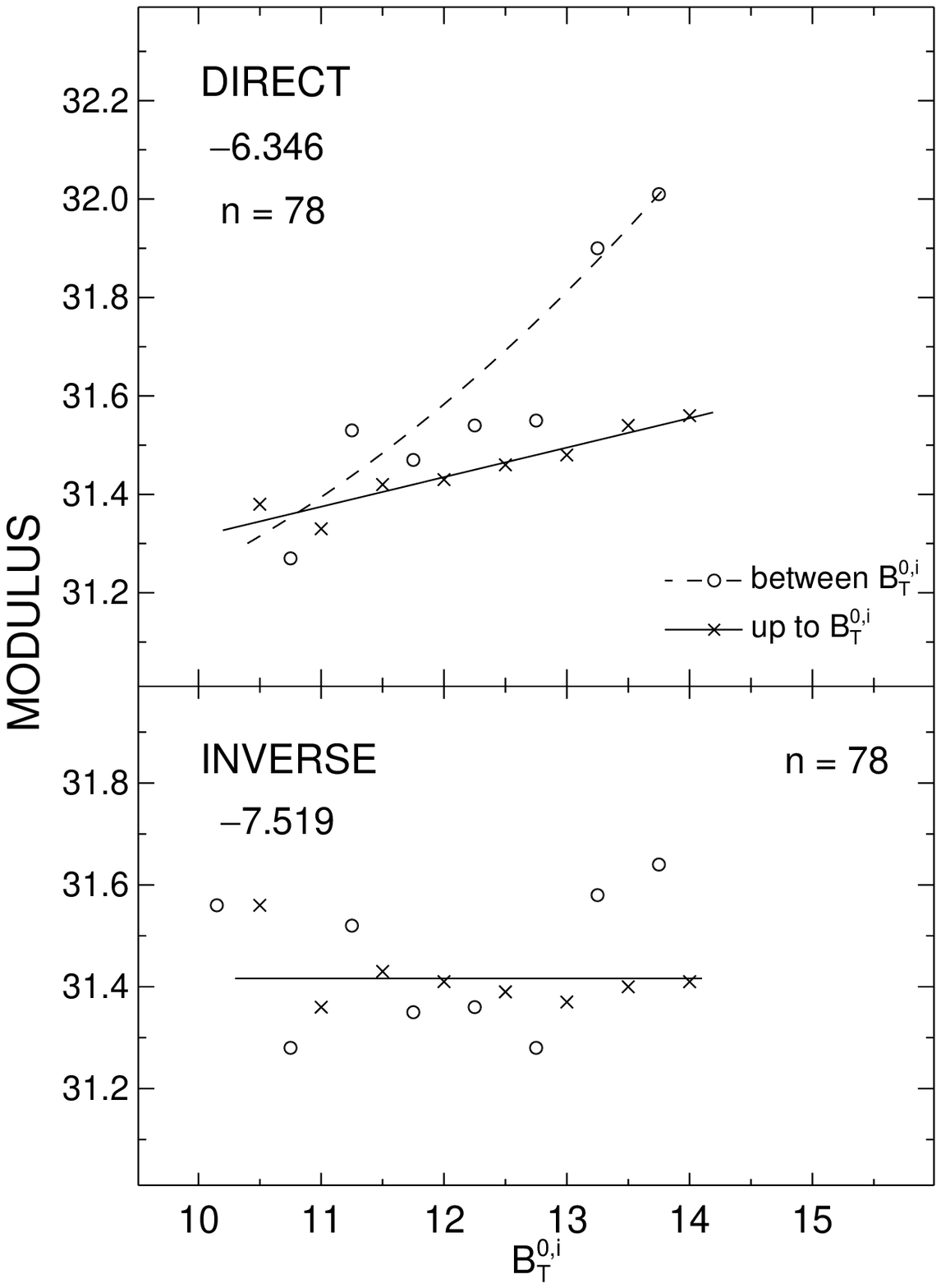}
   \caption{Similar to Figure~\ref{fig:10} but for the restricted
   sample with  $n=78$. The Teerikorpi cluster population
   incompleteness bias is present using the direct TF slope but is
   absent using the inverse slope. The open circles are the modulus
   values from the data summed over all LW and averaged within the
   individual magnitude intervals (Table~\ref{tab:08}A). The crosses
   are the modulus values for the data summed aver all LW and then
   summed to the various magnitude cut-off limits
   (Table~\ref{tab:08}B).} 
\label{fig:11}
\end{figure}
\clearpage


\begin{deluxetable}{cccccc}
\tablewidth{0pt}
\tabletypesize{\footnotesize}
\tablecaption{Derived Distance Moduli of the Toy Cluster of
  Figure~\ref{fig:01} as a Function of $B_{T}^{0,i}$\label{tab:01}}   
\tablehead{
 \colhead{} &
\multicolumn{5}{c}{TF slope}
\\
 \colhead{$B_{T}^{0,i}$}   & 
 \colhead{$-9$} &
 \colhead{$-8$} &
 \colhead{$-7$} &
 \colhead{$-6$} &
 \colhead{$-5$}      
\\
 \colhead{(1)}     & 
 \colhead{(2)}     & 
 \colhead{(3)}     & 
 \colhead{(4)}     & 
 \colhead{(5)}     & 
 \colhead{(6)}     
\\[2pt]
\tableline\\[-8pt]
& \multicolumn{5}{c}{$(m-M)$ at $B_{T}^{0,i}$} 
}
\startdata
10.0 & 31.93 & 31.70 & 31.50 & 31.29 & 31.07 \\
10.5 & 31.79 & 31.64 & 31.50 & 31.36 & 31.21 \\
11.0 & 31.64 & 31.57 & 31.50 & 31.43 & 31.36 \\
11.5 & 31.50 & 31.50 & 31.50 & 31.50 & 31.50 \\
12.0 & 31.36 & 31.43 & 31.50 & 31.57 & 31.64 \\
12.5 & 31.21 & 31.36 & 31.50 & 31.64 & 31.79 \\
13.0 & 31.07 & 31.29 & 31.50 & 31.71 & 31.93 \\
13.5 & 30.93 & 31.21 & 31.50 & 31.79 & 32.07 \\
14.0 & 30.78 & 31.14 & 31.50 & 31.86 & 32.21 \\
14.5 & 30.64 & 31.07 & 31.50 & 31.93 & 32.36 \\
\tableline\\[-8pt]
 \colhead{} &
\multicolumn{5}{c}{$\langle m-M\rangle$ from data summed up to $B_{T}^{0,i}$}
\\[2pt]
\tableline\\[-8pt]
10.0 & 31.93 & 31.70 & 31.50 & 31.29 & 31.07 \\
10.5 & 31.86 & 31.67 & 31.50 & 31.33 & 31.14 \\
11.0 & 31.79 & 31.64 & 31.50 & 31.36 & 31.21 \\
11.5 & 31.72 & 31.60 & 31.50 & 31.40 & 31.28 \\
12.0 & 31.64 & 31.57 & 31.50 & 31.43 & 31.36 \\
12.5 & 31.57 & 31.53 & 31.50 & 31.47 & 31.43 \\
13.0 & 31.50 & 31.50 & 31.50 & 31.50 & 31.50 \\
\tableline
13.5 & 31.43 & 31.46 & 31.50 & 31.54 & 31.57 \\
14.0 & 31.36 & 31.43 & 31.50 & 31.57 & 31.64 \\
14.5 & 31.28 & 31.39 & 31.50 & 31.61 & 31.71 \\
\enddata
\end{deluxetable}

\clearpage

\begin{deluxetable}{cccclcclccc}
\tablewidth{0pt}
\tabletypesize{\footnotesize}
\tablecaption{Least Squares TF Regressions to Various Magnitude 
  Cut-offs for the Total ($n=92$) and the Restricted ($n=78$) Sample of
  the Composite Cluster\label{tab:02}}   
\tablehead{
 \colhead{}                     &
 \colhead{}                     &
 \multicolumn{2}{c}{Total}      &
 \colhead{}                     &
 \multicolumn{2}{c}{Total}      &
 \colhead{}                     &
 \colhead{}                     &
 \multicolumn{2}{c}{Restricted} 
\\
 \colhead{}                     &
 \colhead{}                     &
 \multicolumn{2}{c}{DIRECT}     &
 \colhead{}                     &
 \multicolumn{2}{c}{INV}        &
 \colhead{}                     &
 \colhead{}                     &
 \colhead{DIRECT}               &
 \colhead{INV}                  
\\
\cline{3-4}
\cline{6-7}
 \colhead{$B_{T}^{0,i}$}        & 
 \colhead{$n$}                  &
 \colhead{Slope}                &
 \colhead{Zpt}                  &
 \colhead{}                     &
 \colhead{Slope}                &
 \colhead{Zpt}                  & 
 \colhead{}                     &
 \colhead{$n$}                  &
 \multicolumn{2}{c}{slopes}
\\
 \colhead{(1)}     & 
 \colhead{(2)}     & 
 \colhead{(3)}     & 
 \colhead{(4)}     & 
 \colhead{}        &
 \colhead{(5)}     & 
 \colhead{(6)}     &
 \colhead{}        &
 \colhead{(7)}     & 
 \colhead{(8)}     & 
 \colhead{(9)}     
} 
\startdata
$ < 11.5$ & 28 & $-4.550$ & 22.603 && $-7.645$ & 30.653 && 26 & $-5.023$ & $-7.680$ \\
$ < 12.0$ & 42 & $-4.755$ & 23.290 && $-8.157$ & 31.998 && 38 & $-5.247$ & $-7.158$ \\
$ < 12.5$ & 56 & $-5.106$ & 24.251 && $-7.353$ & 29.909 && 51 & $-5.606$ & $-7.052$ \\
$ < 13.0$ & 70 & $-5.228$ & 24.610 && $-6.935$ & 28.831 && 64 & $-5.720$ & $-6.930$ \\
$ < 13.5$ & 80 & $-5.716$ & 25.888 && $-7.386$ & 29.983 && 74 & $-6.155$ & $-7.364$ \\
$ < 14.0$ & 84 & $-5.936$ & 26.454 && $-7.547$ & 30.391 && 78 & $-6.346$ & $-7.519$ \\
$ < 14.5$ & 87 & $-6.235$ & 27.216 && $-7.831$ & 31.101 && \nodata & \nodata & \nodata \\
$ < 15.0$ & 92 & $-6.622$ & 28.195 && $-8.032$ & 31.603 && \nodata & \nodata & \nodata \\
\enddata
\end{deluxetable}

\clearpage

\begin{deluxetable}{llcrcc}
\tablewidth{0pt}
\tabletypesize{\footnotesize}
\tablecaption{The Cepheid Calibrators\label{tab:03}}   
\tablehead{
 \colhead{Name}             &
 \colhead{Type}             &
 \colhead{$(m-M)$}          &
 \colhead{$B_{T}^{0,i}$}    & 
 \colhead{$M_{B(T)}^{0,i}$} & 
 \colhead{$\log w_{20}$} 
\\
 \colhead{}        &
 \colhead{}        &
 \colhead{Saha\,06}  &
 \colhead{FTS\,98}   & 
 \colhead{FTS\,98}   & 
 \colhead{} 
\\
 \colhead{(1)}     & 
 \colhead{(2)}     & 
 \colhead{(3)}     & 
 \colhead{(4)}     & 
 \colhead{(5)}     & 
 \colhead{(6)}     
} 
\startdata
NGC\,224  & SbI-II       & 24.54 &  3.36 & $-21.18$ & 2.728 \\
NGC\,300  & ScII.8       & 26.48 &  8.49 & $-17.99$ & 2.296 \\
NGC\,598  & Sc(s)II-III  & 24.64 &  5.73 & $-18.91$ & 2.357 \\
NGC\,925  & SBc(s)II-III & 29.84 & 10.04 & $-19.80$ & 2.382 \\ 
NGC\,1365 & SBbc(s)I     & 31.46 &  9.94 & $-21.52$ & 2.648 \\
NGC\,1425 & Sb(r)II      & 31.96 & 10.83 & $-21.13$ & 2.582 \\
NGC\,2090 & Sc(s)II      & 30.48 & 11.52 & $-18.96$ & 2.503 \\
NGC\,2403 & Sc(s)III     & 27.43 &  8.38 & $-19.05$ & 2.415 \\
NGC\,2541 & Sc(s)III     & 30.50 & 11.36 & $-19.14$ & 2.330 \\
NGC\,3031 & Sb(r)I-II    & 27.80 &  7.34 & $-20.46$ & 2.667 \\
NGC\,3198 & Sc(s)I-II    & 30.80 & 10.14 & $-20.66$ & 2.508 \\
NGC\,3319 & SB(s)II.4    & 30.74 & 11.15 & $-19.59$ & 2.394 \\
NGC\,3351 & SBb(r)II     & 30.10 & 10.24 & $-19.86$ & 2.538 \\
NGC\,3368 & Sab(s)II     & 30.34 &  9.85 & $-20.49$ & 2.649 \\
NGC\,3621 & Sc(s)II.8    & 29.30 &  9.40 & $-19.90$ & 2.509 \\
NGC\,3627 & Sb(s)II.2    & 30.50 &  9.07 & $-21.43$ & 2.591 \\
NGC\,4258 & Sb(s)II      & 29.50 &  8.31 & $-21.13$ & 2.649 \\
NGC\,4321 & Sc(s)I       & 31.18 &  9.88 & $-21.30$ & 2.725 \\
NGC\,4535 & SBc(s)I.3    & 31.25 & 10.28 & $-20.97$ & 2.591 \\
NGC\,4536 & Sbc(s)I-II   & 31.24 & 10.62 & $-20.62$ & 2.548 \\    
NGC\,4548 & SBb(rs)I-II  & 30.99 & 10.77 & $-20.22$ & 2.631 \\
NGC\,4639 & SB(r)II      & 32.20 & 11.95 & $-20.25$ & 2.617 \\ 
NGC\,4725 & Sb/SBb(r)II  & 30.65 &  9.73 & $-20.92$ & 2.723 \\
NGC\,5457 & Sc(s)I       & 29.17 &  8.31 & $-20.86$ & 2.665 \\
NGC\,7331 & Sb(rs)I-II   & 30.89 &  9.39 & $-21.50$ & 2.725 \\
\enddata
\end{deluxetable}
\setlength\textheight{9.0in}%

\clearpage

\begin{deluxetable}{rlllllll}
\tablewidth{0pt}
\tabletypesize{\footnotesize}
\tablecaption{$\langle m-M\rangle$ Averaged within Various Discrete
  Intervals of Line Width and Listed in Magnitude Intervals for the
  Composite Cluster\label{tab:04}}   
\tablehead{
 \colhead{}                 &
\multicolumn{7}{c}{Log Line Width Intervals ($w_{20}$)}
\\
 \colhead{$B_{T}^{0,i}$}    & 
 \colhead{2.7-2.8}          &
 \colhead{2.6-2.7}          &
 \colhead{2.5-2.6}          &
 \colhead{2.4-2.5}          &
 \colhead{2.3-2.4}          &
 \colhead{2.2-2.3}          &
 \colhead{2.1-2.2}          
\\
 \colhead{(1)}     & 
 \colhead{(2)}     & 
 \colhead{(3)}     & 
 \colhead{(4)}     & 
 \colhead{(5)}     & 
 \colhead{(6)}     &
 \colhead{(7)}     & 
 \colhead{(8)}     
\\[2pt]
\tableline\\[-8pt]
\multicolumn{8}{c}{(A)  DIRECT SLOPE $=-6.622$}
} 
\startdata
 9.8-10.5 & 31.60\,(5) & 31.13\,(3) & 30.52\,(1) &            &            &            &            \\
10.5-11.0 &            & 31.68\,(3) & 30.94\,(4) &            &            &            &            \\
11.0-11.5 &            & 31.80\,(2) & 30.57\,(7) & 31.10\,(2) & 30.55\,(1) &            &            \\  
11.5-12.0 &            & 32.58\,(2) & 31.90\,(4) & 31.35\,(6) & 30.79\,(2) &            &            \\ 
12.0-12.5 &            &            &            & 31.76\,(7) & 31.22\,(6) & 30.55\,(1) &            \\ 
12.5-13.0 &            &            &            & 32.15\,(2) & 31.74\,(5) & 31.05\,(6) & 30.29\,(9) \\
13.0-13.5 &            &            &            &            & 32.12\,(5) & 31.68\,(4) & 30.90\,(1) \\
13.5-14.0 &            &            &            &            & 32.50\,(1) & 31.79\,(2) & 31.61\,(1) \\
14.0-15.0 &            &            &            &            &            & 32.82\,(1) & 32.12\,(7) \\
Summed    &            &            &            &            &            &            &            \\
\multicolumn{1}{l}{vertical}  & 31.60      & 31.72      & 31.43      & 31.58      & 31.65      & 31.42      & 31.77      \\
  \multicolumn{1}{c}{$n$} & \multicolumn{1}{c}{$5\quad$} &
  \multicolumn{1}{c}{$10\quad$} & \multicolumn{1}{c}{$16\quad$} &
  \multicolumn{1}{c}{$17\quad$} & \multicolumn{1}{c}{$20\quad$} &
  \multicolumn{1}{c}{$14\quad$} & \multicolumn{1}{c}{$10\quad$} \\
\multicolumn{1}{l}{rms} & 0.145 & 0.563 & 0.483   & 0.372      & 0.556      & 0.573      & 0.667      \\
\tableline\\[-8pt]
\multicolumn{8}{c}{(B) INVERSE SLOPE $=-8.032$} \\[2pt]
\tableline\\[-8pt]
 9.8-10.5 &  31.85\,(5) & 31.29\,(3) & 30.56\,(1) &            &            &            &            \\
10.5-11.0 &             & 31.79\,(3) & 30.93\,(4) &            &            &            &            \\
11.0-11.5 &             & 31.88\,(2) & 31.32\,(7) & 30.98\,(2) & 30.29\,(1) &            &            \\ 
11.5-12.0 &             & 32.68\,(2) & 31.86\,(4) & 31.36\,(6) & 30.51\,(2) &            &            \\ 
12.0-12.5 &             &            &            & 31.60\,(6) & 30.13\,(1) &            &            \\
12.5-13.0 &             &            &            & 31.99\,(2) & 31.47\,(5) & 30.61\,(6) & 29.69\,(1) \\ 
13.0-13.5 &             &            &            &            & 31.84\,(5) & 31.26\,(4) & 30.33\,(1) \\ 
13.5-14.0 &             &            &            &            & 32.19\,(1) & 31.34\,(2) & 31.05\,(1) \\
14.0-15.0 &             &            &            &            &            & 32.41\,(1) & 31.54\,(7) \\ 
Summed    &             &            &            &            &            &            &            \\
\multicolumn{1}{l}{vertical} & 31.85 & 31.84 & 31.41 & 31.43   & 31.27      & 30.99      & 31.18      \\
  \multicolumn{1}{c}{$n$} & \multicolumn{1}{c}{$5\quad$} &
  \multicolumn{1}{c}{$10\quad$} & \multicolumn{1}{c}{$16\quad$} &
  \multicolumn{1}{c}{$17\quad$} & \multicolumn{1}{c}{$20\quad$} &
  \multicolumn{1}{c}{$14\quad$} & \multicolumn{1}{c}{$10\quad$} \\
\multicolumn{1}{l}{rms} & 0.179 & 0.552 & 0.481   & 0.389      & 0.561      & 0.580      & 0.673      \\
\enddata
\end{deluxetable}
\setlength\textheight{8.4in}%

\clearpage

\begin{deluxetable}{rrccc@{$\qquad\qquad$}crcc}
\tablewidth{0pt}
\tabletypesize{\footnotesize}
\tablecaption{$\langle m-M\rangle$ as Summed over all Line Widths for
  Various Magnitude Intervals and Cut-off Magnitudes for the Composite
  Cluster\label{tab:05}}    
\tablehead{
 \colhead{between}                  & 
 \colhead{}                         & 
 \multicolumn{2}{c}{summed over LW} & 
 \colhead{}                         & 
 \colhead{up to}                    & 
 \colhead{}                         & 
 \multicolumn{2}{c}{summed over LW}  
\\
\cline{3-4}
\cline{8-9}
 \colhead{$B_{T}^{0,i}$}        & 
 \colhead{$n$}                  &
 \colhead{$\langle m-M\rangle$} &
 \colhead{rms}                  &
 \colhead{}                     & 
 \colhead{$B_{T}^{0,i}$}        &
 \colhead{$n$}                  &
 \colhead{$\langle m-M\rangle$} &
 \colhead{rms}                  
\\
 \colhead{(1)}     & 
 \colhead{(2)}     & 
 \colhead{(3)}     & 
 \colhead{(4)}     & 
 \colhead{}        & 
 \colhead{(5)}     & 
 \colhead{(6)}     &
 \colhead{(7)}     & 
 \colhead{(8)}     
} 
\startdata
\multicolumn{9}{c}{(A)  DIRECT SLOPE $=-6.622$}\\[2pt]
\tableline
 9.8-10.5 &  9 & 31.32 & 0.398 && 10.5 &  9 & 31.32 & 0.398 \\
10.5-11.0 &  7 & 31.25 & 0.470 && 11.0 & 16 & 31.29 & 0.417 \\
11.0-11.5 & 12 & 31.45 & 0.410 && 11.5 & 28 & 31.36 & 0.414 \\ 
11.5-12.0 & 14 & 31.60 & 0.570 && 12.0 & 42 & 31.44 & 0.479 \\
12.0-12.5 & 14 & 31.44 & 0.427 && 12.5 & 56 & 31.44 & 0.463 \\
12.5-13.0 & 14 & 31.40 & 0.550 && 13.0 & 70 & 31.43 & 0.478 \\
13.0-13.5 & 10 & 31.82 & 0.402 && 13.5 & 80 & 31.48 & 0.484 \\
13.5-14.0 &  4 & 31.92 & 0.413 && 14.0 & 84 & 31.50 & 0.488 \\
14.0-15.0 &  8 & 32.21 & 0.286 && 15.0 & 92 & 31.56 & 0.514 \\
\tableline\\[-8pt]
\multicolumn{9}{c}{(B) INVERSE SLOPE $=-8.032$}\\[2pt]
\tableline
 9.8-10.5 &  9 & 31.52 & 0.475 && 10.5 &  9 & 31.52 & 0.475 \\
10.5-11.0 &  7 & 31.30 & 0.550 && 11.0 & 16 & 31.42 & 0.504 \\
11.0-11.5 & 12 & 31.40 & 0.510 && 11.5 & 28 & 31.42 & 0.495 \\
11.5-12.0 & 14 & 31.50 & 0.693 && 12.0 & 42 & 31.44 & 0.561 \\
12.0-12.5 & 14 & 31.20 & 0.527 && 12.5 & 56 & 31.38 & 0.558 \\
12.5-13.0 & 14 & 31.05 & 0.676 && 13.0 & 70 & 31.32 & 0.594 \\
13.0-13.5 & 10 & 31.44 & 0.495 && 13.5 & 80 & 31.33 & 0.581 \\
13.5-14.0 &  4 & 31.48 & 0.300 && 14.0 & 84 & 31.34 & 0.576 \\
14.0-15.0 &  8 & 31.65 & 0.354 && 15.0 & 92 & 31.36 & 0.566 \\
\enddata
\end{deluxetable}

\clearpage

\begin{deluxetable}{crcclccrlcclcc}
\tablewidth{0pt}
\tabletypesize{\footnotesize}
\tablecaption{$\langle m-M\rangle$ as Summed to Various Magnitude
  Cut-offs over all LW for Different Tully-Fisher Slopes\label{tab:06}}    
\tablehead{
 \multicolumn{2}{c}{Summed to} & 
 \multicolumn{11}{c}{SLOPES}  
\\
 \colhead{}                    & 
 \colhead{}                    & 
 \multicolumn{2}{c}{$-9$}      & 
 \colhead{}                    & 
 \multicolumn{2}{c}{$-7$}      & 
 \colhead{}                    & 
 \multicolumn{2}{c}{$-6$}      & 
 \colhead{}                    & 
 \multicolumn{2}{c}{$-5$} 
\\
\cline{3-4}
\cline{6-7}
\cline{9-10}
\cline{12-13}
 \colhead{$B_{T}^{0,i}$}        & 
 \colhead{$n$}                  &
 \colhead{$\langle m-M\rangle$} &
 \colhead{rms}                  &
 \colhead{}                     & 
 \colhead{$\langle m-M\rangle$} &
 \colhead{rms}                  &
 \colhead{}                     & 
 \colhead{$\langle m-M\rangle$} &
 \colhead{rms}                  &
 \colhead{}                     & 
 \colhead{$\langle m-M\rangle$} &
 \colhead{rms}                  
\\
 \colhead{(1)}     & 
 \colhead{(2)}     & 
 \colhead{(3)}     & 
 \colhead{(4)}     & 
 \colhead{}        & 
 \colhead{(5)}     & 
 \colhead{(6)}     &
 \colhead{}        & 
 \colhead{(7)}     & 
 \colhead{(8)}     &
 \colhead{}        & 
 \colhead{(9)}     & 
 \colhead{(10)}     
} 
\startdata
10.5 &  9 & 31.66 & .524 && 31.37 & .418 && 31.23 & .365 && 31.09 & .317 \\
11.0 & 16 & 31.51 & .567 && 31.33 & .439 && 31.23 & .383 && 31.14 & .339 \\
11.5 & 28 & 31.45 & .563 && 31.37 & .433 && 31.33 & .388 && 31.29 & .366 \\ 
12.0 & 42 & 31.44 & .632 && 31.44 & .498 && 31.44 & .455 && 31.44 & .436 \\
12.5 & 56 & 31.34 & .643 && 31.43 & .485 && 31.46 & .437 && 31.50 & .423 \\
13.0 & 70 & 31.23 & .697 && 31.40 & .504 && 31.48 & .447 && 31.56 & .434 \\
13.5 & 80 & 31.23 & .678 && 31.44 & .504 && 31.55 & .467 && 31.65 & .497 \\ 
14.0 & 84 & 31.22 & .671 && 31.46 & .505 && 31.57 & .477 && 31.69 & .497 \\
15.0 & 92 & 31.23 & .651 && 31.51 & .518 && 31.65 & .525 && 31.78 & .576 \\
\enddata
\end{deluxetable}

\clearpage

\begin{deluxetable}{cccccccc}
\tablewidth{0pt}
\tabletypesize{\footnotesize}
\tablecaption{Regressions for the Cepheid Calibration and for the
  Composite Cluster for Various Tully-Fisher Slopes: 
  the Sensitivity of the Resulting Distance Moduli to Slope\label{tab:07}}   
\tablehead{
 \colhead{Slope} &
 \multicolumn{2}{c}{Cepheid calib.\tablenotemark{5)}} & 
 \multicolumn{2}{c}{Cluster\tablenotemark{6)}} & 
 \colhead{n} &
 \colhead{$\langle m-M\rangle$} &
 \colhead{$B_{T}^{0,i}$}
\\
 \colhead{a}   & 
 \colhead{b}   &
 \colhead{rms} &
 \colhead{c}   &
 \colhead{rms} &
 \colhead{}    &
 \multicolumn{2}{c}{to sample end}
\\
 \colhead{(1)}     & 
 \colhead{(2)}     & 
 \colhead{(3)}     & 
 \colhead{(4)}     & 
 \colhead{(5)}     & 
 \colhead{(6)}     & 
 \colhead{(7)}     & 
 \colhead{(8)}     
} 
\startdata
$-9$                       & $+2.716$ & 0.624 & 33.946 & 0.651 & 92 & 31.23 & 15.0 \\
$-8$                       & $+0.197$ & 0.541 & 31.525 & 0.564 & 92 & 31.33 & 15.0 \\
$-7$                       & $-2.402$ & 0.511 & 29.108 & 0.518 & 92 & 31.51 & 15.0 \\
$-6$                       & $-4.961$ & 0.499 & 26.689 & 0.525 & 92 & 31.65 & 15.0 \\
$-5$                       & $-7.519$ & 0.522 & 24.261 & 0.576 & 92 & 31.78 & 15.0 \\[5pt]
$-6.161$\tablenotemark{1)} & $-4.549$ & 0.500 & 27.078 & 0.520 & 92 & 31.63 & 15.0 \\ 
$-8.475$\tablenotemark{1)} & $+1.364$ & 0.580 & 32.664 & 0.598 & 92 & 31.30 & 15.0 \\[5pt]
$-6.622$\tablenotemark{2)} & $-3.369$ & 0.502 & 28.191 & 0.514 & 92 & 31.56 & 15.0 \\
$-8.032$\tablenotemark{2)} & $+0.239$ & 0.557 & 31.609 & 0.566 & 92 & 31.37 & 15.0 \\[5pt]
$-5.936$\tablenotemark{3)} & $-5.124$ & 0.500 & 26.456 & 0.477 & 84 & 31.58 & 14.0 \\
$-7.547$\tablenotemark{3)} & $-1.003$ & 0.533 & 30.387 & 0.538 & 84 & 31.39 & 14.0 \\[5pt]
$-6.346$\tablenotemark{4)} & $-4.077$ & 0.499 & 27.483 & 0.415 & 78 & 31.56 & 14.0 \\
$-7.519$\tablenotemark{4)} & $-1.077$ & 0.530 & 30.343 & 0.451 & 78 & 31.42 & 14.0 \\
\enddata
\tablenotetext{1)}{Cepheid least squares regressions, n = 25}
\tablenotetext{2)}{Direct and inverse slopes for total cluster sample (Fig.~\ref{fig:05})}
\tablenotetext{3)}{Direct and inverse slopes for cluster sample cut at
   $B_{T}^{0,i}=14.0$ (Fig.~\ref{fig:06})}
\tablenotetext{4)}{Direct and inverse slopes for final sample (with a $B = 14.0$ 
cut plus six others eliminated)}
\tablenotetext{5)}{Cepheid calibration as $M_{B(T)}  = a \log w_{20}+b$}
\tablenotetext{6)}{Cluster regression as $B_{T}^{0,i}= a \log w_{20}+c$}
\end{deluxetable}

\clearpage

\begin{deluxetable}{rrlcclcc}
\tablewidth{0pt}
\tabletypesize{\footnotesize}
\tablecaption{Modulus of the Composite Cluster for Various Magnitude
  Intervals and Limits from the High Weight Restricted Sample\label{tab:08}}  
\tablehead{
 \colhead{}                         &  
 \colhead{}                         & 
 \colhead{}                         & 
 \multicolumn{2}{c}{Slope$=-6.346$} & 
 \colhead{}                         & 
 \multicolumn{2}{c}{Slope$=-7.519$} 
\\
 \colhead{Interval}                 & 
 \colhead{}                         & 
 \colhead{}                         & 
 \multicolumn{2}{c}{Direct (78)}    & 
 \colhead{}                         & 
 \multicolumn{2}{c}{Inverse (78)} 
\\
\cline{4-5}
\cline{7-8}
 \colhead{$B_{T}^{0,i}$}        &
 \colhead{$n$}                  &
 \colhead{}                     & 
 \colhead{$\langle m-M\rangle$} &
 \colhead{rms}                  &
 \colhead{}                     & 
 \colhead{$\langle m-M\rangle$} &
 \colhead{rms}                  
\\
 \colhead{(1)}     & 
 \colhead{(2)}     & 
 \colhead{}        & 
 \colhead{(3)}     & 
 \colhead{(4)}     & 
 \colhead{}        & 
 \colhead{(5)}     & 
 \colhead{(6)}
\\[2pt]
\tableline\\[-8pt]
\multicolumn{8}{c}{(A) IN MAGNITUDE INTERVALS}
} 
\startdata
 9.8-10.5 &  8  &&  31.78 & .270  &&  31.56 & .311 \\
10.5-11.0 &  7  &&  31.27 & .505  &&  31.28 & .521 \\
11.0-11.5 & 11  &&  31.53 & .299  &&  31.52 & .355 \\    
11.5-12.0 & 12  &&  31.47 & .407  &&  31.35 & .482 \\  
12.0-12.5 & 13  &&  31.54 & .339  &&  31.36 & .412 \\
12.5-13.0 & 13  &&  31.55 & .446  &&  31.28 & .535 \\   
13.0-13.5 & 10  &&  31.90 & .383  &&  31.58 & .461 \\
13.5-14.0 &  4  &&  32.01 & .396  &&  31.64 & .480 \\
\tableline\\[-8pt]
\multicolumn{8}{c}{(B) TO MAGNITUDE CUT-OFF LIMITS}\\[2pt]
\tableline\\[-8pt]
     10.5 &  8  &&  31.38 & .270  &&  31.56 & .311 \\ 
     11.0 & 15  &&  31.33 & .386  &&  31.36 & .442 \\
     11.5 & 26  &&  31.42 & .360  &&  31.43 & .407 \\  
     12.0 & 38  &&  31.43 & .371  &&  31.41 & .427 \\
     12.5 & 51  &&  31.46 & .363  &&  31.39 & .420 \\ 
     13.0 & 64  &&  31.48 & .379  &&  31.37 & .444 \\ 
     13.5 & 74  &&  31.54 & .404  &&  31.40 & .449 \\ 
     14.0 & 78  &&  31.56 & .415  &&  31.41 & .451 \\ 
\enddata
\end{deluxetable}


\begin{thebibliography}{}
%
\bibitem[Binggeli et~al.(1993)Binggeli, Popescu, \& Tammann]{Binggeli:etal:93}
   Binggeli, B., Popescu, C.~C., \& Tammann, G.~A. 1993, 
   A\&AS, 98, 275
%
\bibitem[Binggeli et~al.(1985)Binggeli, Sandage, \& Tammann]{Binggeli:etal:85}
   Binggeli, B., Sandage, A., \& Tammann, G.~A. 1985, 
   AJ, 90, 1681
%
\bibitem[Bottinelli et~al.(1971)]{Bottinelli:etal:71}
   Bottinelli, L., Chamaraux, P., G{\'e}rard, E., Gouguenheim, L., 
   Heidmann, J., Kaz{\`e}s, I., \& Lauqu{\'e}, R. 1971, 
   A\&A, 12, 264
%
\bibitem[Bottinelli et~al.(1987)]{Bottinelli:etal:87}
   Bottinelli, L., Gouguenheim, L., Fouqu{\'e}, P., Paturel, G., \& 
   Teerikorpi, P. 1987, 
   A\&A, 181, 1
%
\bibitem[Bottinelli et~al.(1986)]{Bottinelli:etal:86}
   Bottinelli, L., Gouguenheim, L., Paturel, G., \& Teerikorpi, P. 1986, 
   A\&A, 156, 157
%
\bibitem[Bottinelli et~al.(1988a)]{Bottinelli:etal:88a}
   Bottinelli, L., Gouguenheim, L., Paturel, G., \& Teerikorpi, P. 1988a,
   ApJ, 328, 4
%
\bibitem[Bottinelli et~al.(1988b)Bottinelli, Gouguenheim, \& Teerikorpi]{Bottinelli:etal:88b}
   Bottinelli, L., Gouguenheim, L., \& Teerikorpi, P. 1988b, 
   A\&A, 196, 17
%
\bibitem[de Vaucouleurs(1975)]{deVaucouleurs:75}
   de~Vaucouleurs, G. 1975, 
   in Galaxies and the Universe, eds. 
   A.~Sandage, M.~Sandage, \& J.~Kristian, Vol. 9 of Univ. Chicago 
   Press Compendium, 557 
%
\bibitem[de Vaucouleurs et~al.(1991)]{deVaucouleurs:etal:91}
   de~Vaucouleurs, G., de~Vaucouleurs, A., Corwin, H.~G., Buta, R.~J., 
   Paturel, G., \& Fouqu{\'e}, P. 1991, 
   Third Catalogue of Bright Galaxies (New York: Springer) (RC3)
%
\bibitem[Ekholm \& Teerikorpi(1997)]{Ekholm:Teerikorpi:97}
   Ekholm, T., \& Teerikorpi, P. 1997, 
   A\&A, 325, 33
%
\bibitem[Federspiel(1999)]{Federspiel:99}
   Federspiel, M. 1999, 
   PhD thesis, University Basel
%
\bibitem[Federspiel, Sandage \& Tammann(1994)FST\,94]{FST:94}
   Federspiel, M., Sandage, A., \& Tammann, G.~A. 1994, 
   ApJ, 430, 29 (Paper III), (FST\,94)
%
\bibitem[Federspiel, Tammann, \& Sandage(1998)FTS\,98]{FTS:98}
   Federspiel, M., Tammann, G.~A., \& Sandage, A. 1998, 
   ApJ, 495, 115 (FTS\,98)
%
\bibitem[Feigelson \& Babu(1992)]{Feigelson:Babu:92}
   Feigelson, E.~D., \& Babu, G.~J. 1992, 
   ApJ, 397, 55
%
\bibitem[Fouqu{\'e} et~al.(1990)]{Fouque:etal:90}
   Fouqu{\'e}, P., Bottinelli, L., Gouguenheim, L., \& Paturel, G. 1990, 
   ApJ, 349, 1 
%
\bibitem[Freedman et~al.(2001)]{Freedman:etal:01}
   Freedman, W.~L., et~al. 2001, 
   ApJ, 553, 47
%
\bibitem[Giovanelli et~al.(1997a)]{Giovanelli:etal:97a}
   Giovanelli, R., Haynes, M.~P., Herter, T., Vogt, N.~P., 
   Wegner, G., Salzer, J.~J., da~Costa, L.~N., \& Freudling, W. 1997a, 
   AJ, 113, 22
%
\bibitem[Giovanelli et~al.(1997b)]{Giovanelli:etal:97b}
   Giovanelli, R., et~al. 1997b,
   AJ, 113, 53
%
\bibitem[Gouguenheim(1969)]{Gouguenheim:69}
   Gouguenheim, L. 1969,
   A\&A, 3, 281
%
\bibitem[Hendry \& Simmons(1994)]{Hendry:Simmons:94}
   Hendry, M.~A., \& Simmons, J.~F.~L. 1994,
   Vistas Astron., 39, 297
%
\bibitem[Humason et~al.(1956)Humason, Mayall, \& Sandage]{Humason:etal:56}
   Humason, M.~L., Mayall, N.~U., \& Sandage, A.~R. 1956, 
   AJ, 61, 97 
%
\bibitem[Jerjen \& Tammann(1993)]{Jerjen:Tammann:93}
   Jerjen, H., \& Tammann, G.~A. 1993, 
   A\&A, 276, 1
%
\bibitem[Kanbur \& Ngeow(2004)]{Kanbur:Ngeow:04}
   Kanbur, S.~M., \& Ngeow, C.~C. 2004, 
   MNRAS, 350, 962 
%
\bibitem[Kapteyn(1914)]{Kapteyn:14}
   Kapteyn, J.~C. 1914, 
   ApJ, 40, 43 
%
\bibitem[Kraan-Korteweg(1986)]{Kraan-Korteweg:86}
   Kraan-Korteweg, R.~C. 1986, 
   A\&AS, 66, 255
%
\bibitem[Kraan-Korteweg et~al.(1988)Kraan-Korteweg, Cameron, \& Tammann]{Kraan-Korteweg:etal:88}
   Kraan-Korteweg, R.~C., Cameron, L.~M., \& Tammann, G.~A. 1988, 
   ApJ, 331, 620
%
\bibitem[Klypin et~al.(2003)]{Klypin:etal:03}
   Klypin, A., Hoffman, Y., Kravtsov, A.~V., \& Gottl{\"o}ber, S. 2003, 
   ApJ, 596, 19
%
\bibitem[Malmquist(1920)]{Malmquist:20}
   Malmquist, K.~G. 1920, 
   %
   Lund Medd. Ser.~II, 22, 1
%
\bibitem[Malmquist(1922)]{Malmquist:22}
   Malmquist, K.~G. 1922,
   %
   Lund Medd. Ser.~I, 100, 1
%
\bibitem[Masters et~al.(2006)]{Masters:etal:06}
   Masters, K.~L., Springob, C.~M., Haynes, M.~P., \& Giovanelli, R. 2006,
   ApJ, 653, 861
%
\bibitem[Ngeow \& Kanbur(2005)]{Ngeow:Kanbur:05}
   Ngeow, C.~C., \& Kanbur, S.~M. 2005, 
   MNRAS, 360, 1033
%
\bibitem[Ngeow et~al.(2005)]{Ngeow:etal:05}
   Ngeow, C.~C, Kanbur, S.~M., Nikolaev, S., Buonaccorsi, J., Cook, 
   K.~H., \& Welch, D.~L. 2005, 
   MNRAS, 363, 831
%
\bibitem[Peletier \& Willner(1993)]{Peletier:Willner:93}
   Peletier, R.~F., \& Willner, S.~P. 1993, 
   ApJ, 418, 626
%
\bibitem[Pierce \& Tully(1988)]{Pierce:Tully:88}
   Pierce, M.~J., \& Tully, R.~B. 1988, 
   ApJ, 330, 579
%
\bibitem[Roberts(1969)]{Roberts:69}
   Roberts, M.~S. 1969, 
   AJ, 74, 859
%
\bibitem[Saha et~al.(2006)]{Saha:etal:06}
   Saha, A., Thim, F., Tammann, G.~A., Reindl, B., \& Sandage, A. 2006,
   ApJS, 165, 108
%
\bibitem[Sandage(1988a)]{Sandage:88a}
   Sandage, A. 1988a, 
   ApJ, 331, 583
%
\bibitem[Sandage(1988b)]{Sandage:88b}
   Sandage, A. 1988b,
   ApJ, 331, 605
%
\bibitem[Sandage(1994a)]{Sandage:94a}
   Sandage, A. 1994a,
   ApJ, 430, 1 (Paper~I)
%
\bibitem[Sandage(1994b)]{Sandage:94b}
   Sandage, A. 1994b,
   ApJ, 430, 13 (Paper~II)
%
\bibitem[Sandage(1999)]{Sandage:99}
   Sandage, A. 1999,
   AJ, 117, 157 (Paper~VII)
%
\bibitem[Sandage \& Tammann(1981, 1987)]{Sandage:Tammann:87}
   Sandage, A., \& Tammann, G.~A. 1981, 1987,
   A Revised Shapley Ames Catalog of Bright Galaxies, 
   Carnegie Pub. 635 (Washington: Carnegie Institution of Washington) 
%
\bibitem[Sandage, Tammann, \& Federspiel(1995)STF\,95]{STF:95}
   Sandage, A., Tammann, G.~A., \& Federspiel, M. 1995,
   ApJ, 452, 1 (Paper~IV), (STF\,95)
%
\bibitem[Sandage et~al.(2004)Sandage, Tammann, \& Reindl]{STR:04}
   Sandage, A., Tammann, G.~A., \& Reindl, B. 2004, 
   A\&A, 424, 43
%
\bibitem[Sandage et~al.(2006)]{Sandage:etal:06}
   Sandage, A., Tammann, G.~A., Saha, A., Reindl, B., Macchetto, F.~D., 
   \& Panagia, N. 2006,
   ApJ, 653, 843
%
\bibitem[Schechter(1980)]{Schechter:80}
   Schechter, P.~L. 1980,
   AJ, 85, 801
%
\bibitem[Scott(1957)]{Scott:57}
   Scott, E.~L. 1957,
   AJ, 62, 248
%
\bibitem[Seares(1944)]{Seares:44}
   Seares, F.~H. 1944,
   ApJ, 100, 255
%
\bibitem[Sersic(1960)]{Sersic:60}
   S{\'e}rsic, J.~L. 1960, 
   Zs. f. Ap, 50, 168
%
\bibitem[Tammann et~al.(2003)Tammann, Sandage, \& Reindl]{TSR:03}
   Tammann, G.~A., Sandage, A., \& Reindl B. 2003, 
   A\&A, 404, 423 
%
\bibitem[Tammann et~al.(2007)Tammann, Sandage, \& Reindl]{TSR:07}
   Tammann, G.~A., Sandage, A., \& Reindl, B. 2007, 
   ApJ, submitted 
%
\bibitem[Teerikorpi(1984)]{Teerikorpi:84}
   Teerikorpi, P. 1984,
   A\&A, 141, 407
%
\bibitem[Teerikorpi(1987)]{Teerikorpi:87}
   Teerikorpi, P. 1987,
   A\&A, 173, 39
%
\bibitem[Teerikorpi(1990)]{Teerikorpi:90}
   Teerikorpi, P. 1990,
   A\&A, 234, 1
%
\bibitem[Teerikorpi(1997)]{Teerikorpi:97}
   Teerikorpi, P. 1997,
   ARA\&A, 35, 101
%
\bibitem[Teerikorpi et~al.(1999)]{Teerikorpi:etal:99}
   Teerikorpi, P., Ekholm, T., Hanski, M.~O., \& Theureau, G. 1999, 
   A\&A, 343, 713 
%
\bibitem[Theureau et~al.(1997a)]{Theureau:etal:97a}
   Theureau, G., Hanski, M., Ekholm, T., Bottinelli, L., 
   Gouguenheim, L., Paturel, G., \& Teerikorpi, P. 1997a, 
   A\&A, 322, 730
%
\bibitem[Theureau et~al.(1997b)]{Theureau:etal:97b}
   Theureau, G., Hanski, M., Teerikorpi, P., Bottinelli, L., 
   Ekholm, T., Gouguenheim, L., \& Paturel, G. 1997b, 
   A\&A, 319, 435 
%
\bibitem[Tully(1988)]{Tully:88}
   Tully, R.~B. 1988, 
   Nature, 334, 209
%
\bibitem[Tully \& Fisher(1977)]{Tully:Fisher:77}
   Tully, R.~B., \& Fisher, J.~R. 1977,
   A\&A, 54, 661
%
\bibitem[Tully \& Fisher(1987)]{Tully:Fisher:87}
   Tully, R.~B., \& Fisher, J.~R. 1987, 
   Atlas of Nearby Galaxies (Cambridge: Cambridge Univ. Press)   
%
\bibitem[Tully et~al.(1996)]{Tully:etal:96}
   Tully, R.~B., Verheijen, M.~A.~W., Pierce, M.~J., Huang, J.-S., \& 
   Wainscoat, R.~J. 1996, 
   AJ, 112, 2471
%
\bibitem[Verheijen(1997)]{Verheijen:97}
   Verheijen, M.~A.~W. 1997,
   PhD thesis, University Groningen
%
\bibitem[Willick(1994)]{Willick:94}
   Willick, J.~A. 1994, 
   ApJS, 92, 1
%
%
\bibitem[Yahil et~al.(1977)Yahil, Tammann, \& Sandage]{Yahil:etal:77}
   Yahil, A., Tammann, G.~A., \& Sandage, A. 1977, 
   ApJ, 217, 903


\end{thebibliography}
\end{document}